\documentclass[onecolumn]{aa}

\newcommand{\greq}{$\stackrel{>}{ _{\sim}}$}
\newcommand{\lteq}{$\stackrel{<}{ _{\sim}}$}

\usepackage{graphicx}
\begin{document}
   \title{Differentially rotating magnetised neutron stars: production of
       toroidal magnetic fields}

   \subtitle{}

   \author{A.V. Thampan
          }

   \offprints{A.V. Thampan}

   \institute{Inter--University Centre for Astronomy and Astrophysics 
              (IUCAA), Pune 411 007, India
	      \email{arun@iucaa.ernet.in}
              }

   \date{}

   \abstract{ We initiate numerical studies of differentially rotating 
   magnetised (proto) neutron stars by studying - through construction 
   from first principles - the coupling between an assumed differential 
   rotation and an impressed magnetic field.  For a perfect incompressible,
   homogeneous, non-dissipative fluid sphere immersed in an ambient
   plasma, we solve the (coupled) azimuthal components of the Navier-Stokes 
   equation and the Maxwell induction equation.  The assumed
   time--independent poloidal field lines get dragged by the rotating fluid 
   and produce toroidal magnetic fields.  Surface magnetic fields take
   away energy redistributing the angular momentum to produce rigid
   rotation along poloidal field lines.  Due to absence of viscous
   dissipation, sustained torsional oscillations are set up within the
   star.  However, the perpetual 
   oscillations of neighbouring `closed' field lines get increasingly out
   of phase with time, leading to structure build up as in Liu \& Shapiro
   (2004) implying the importance of taking into account diffusion (Spruit
   1999) for realistic modeling.

   }

   \maketitle
\section{Introduction}

Absence of short period (\lteq~$1$~ms) newborn radio pulsars (see e.g. 
                        Bhattacharya \& Van den Heuvel 1991) is a long 
standing conundrum of pulsar physics.  Mechanisms for spinning down 
                        the protoneutron star -- the object left 
behind immediately ($3$--$4$~s) after a supernova explosion or a
binary neutron star merger event -- over dynamical timescales, therefore,
assume significance in modeling these objects.  The proposed mechanisms 
of spin--down are through the emission of gravitational waves 
(e.g. Lindblom, Owen \& Morsink 1998), or through magnetic braking or 
viscous damping (Heger et al 2003). Other processes that rely
on instabilities (e.g. Andersson 2003; Spruit \& Phinney 1998) 
for spinning down the star have also received attention.
An important ingredient in all these mechanisms is the differential
rotation of the object.  Differential rotation is expected to be
strong in the first few seconds after the supernova collapse 
(Akiyama et al. 2003) -- this can be the case even following a 
neutron star--neutron star binary merger (Zwerger \& Muller 1997; 
Rampp, Muller \& Ruffert 1998), although such a binary merger results in a
black--hole formation, following a short lived ``hypermassive'' neutron star 
stage (e.g. Shibata \& Uryu 2000, Rasio  \& Shapiro 1999).  Amongst the above 
mechanisms, the spin down of neutron stars through 
gravitational wave emission as a consequence of r--mode oscillations have 
aroused particular interest in recent years (Lindblom, Owen \& Morsink 1998).  
Particularly, there have been suggestions that neutron stars with
superfluid cores and/or solid crusts could be sources for
gravitational waves, spinning down young neutron stars to the observed
pulsar periods (Andersson \& Kokkotas 2001).  The interest in r--mode
instability is due to the possibility of the frequency and amplitude of the 
gravitational waves being in the observable range of upcoming and operational 
gravitational wave detectors like LIGO, VIRGO, TAMA, GEO 
(Heger et al 1998).  However, r--modes can suffer suppression or damping in 
the presence of hyperon bulk viscosity (Jones 2001; there nonetheless
exist ``windows of opportunity'' -- Andersson \& Kokkotas 2001)
or magnetic fields 
(Rezzolla, Lamb \& Shapiro 2000; Rezzolla et al 2001a,b), in addition, it was 
recently shown 
(Arras et al. 2003) that the saturation energies of r--modes are extremely 
small to account for the spin--down to periods of the order of $10$--$100$~ms. 
The damping timescales of differential rotation, varies from $10^{-3}$~s for 
r--modes, to $10^3$~s for Alf\'{v}en effects or $10^9$~s for molecular 
viscosity.  Thus, it is of pertinence for pulsar physics, as well as 
gravitational wave physics, to establish which of the above effects dominate 
and at what timescales they take effect.

Although a magnetic field of $10^{13}$~G is expected by flux conservation 
of the magnetic field of the progenitor star, the exact evolution to a cold 
rigidly rotating object with a predominantly dipole magnetic field 
of the above magnitude is still a controversial issue
(e.g. Chanmugam, Rajasekhar \& Young 1995 and references therein).
The issue is complicated due to observations (for at least one pulsar) 
of higher mulipole components (Gil \& Mitra 2001).
The subsequent evolution of the magnetic field of the radio pulsar 
(Srinivasan et al 1990, Ruderman, M. 1991 a,b,c, Blandford, Applegate \& 
Hernquist, L. 1983, Sang \& Chanmugam 1987, Geppert \& Urpin 1994,
Urpin \& Geppert 1995,1996, Konar \& Bhattacharya 1997,
Konar \& Bhattacharya 1999,
Mitra, Konar \& Bhattacharya 1999, see Bhattacharya 2002 for a recent review), 
thus, crucially depends on the magnetic field
(and the rotation rate) that the pulsar possesses as soon as it is `switched
on' (Vivekanand \& Narayan 1981).  These effects acquire additional
significance when gravitational effects are taken into account (Rezzolla,
Ahmedov \& Miller 2001; Zanotti \& Rezzolla 2002).

A possible mechanism to slow down the pulsar is through the production of
interior toroidal magnetic field components (from poloidal components through
differential rotation) and back reaction of this on the rotational flow.
Moffatt (1978) discussed the distortion of an initially uniform
magnetic field by differential rotation and showed that 
a net toroidal magnetic field flux over
the entire space covering the interior of the star to infinity,
can develop only as a result of diffusion and only if the 
term $\rho(\mathbf{B}_{\rm P} {\bf .} \mathbf{\nabla})\Omega$
(where $\rho$ is the cylindrical radial coordinate, $\mathbf{B}_{\rm P}$ 
the poloidal magnetic field, $\mathbf{\nabla}$ the differential
operator and $\Omega$ the rotation rate) is symmetric about the mid plane. 
Following such an argument, an  analytical expression was obtained for the 
final toroidal field at the asymptotic radial limit for a given constant of 
diffusion.  Mestel (1999) explained the need for toroidal currents to 
maintain the poloidal fields and {\it vice versa} in the interiors of stars 
and presented a wave equation for the rotation rate.  Spruit (1999)  
discussed the interplay between differential rotation and magnetic
fields in stellar interiors, focusing on the inherent 
instabilities that may occur in such systems. 

These works have largely been based on analytical results or analytical 
extensions of the  kinematic dynamo equations. There is a need to
confirm and extend these results using realistic magneto--hydrodynamic 
simulations.  However, such numerical computations have been difficult due
to the inherent non--deterministic nature of the problem and due to the 
several numerical instabilities that can occur during the
simulation. Only recently have reliable and stable numerical schemes
been formulated and initial steps towards  realistic simulations
of these systems been undertaken. Shapiro (2000) solved the 
incompressible, non--turbulent, non--diffusive hydromagnetic dynamo 
equations both analytically and numerically and showed that
for assumed seed magnetic fields, viscosity effectively brakes differential 
rotation.  It was also shown that an infinite cylindrical star, with an 
impressed internal monopolar field, can lose angular momentum to an external 
plasma medium, even in the absence of viscosity.  The effects of 
compressibility was discussed by Cook, Shapiro \& Stephens (2003) who showed 
that radial flow and shocks become important in the evolution.  
Liu \& Shapiro (2004) worked out these effects for relativistic spherical 
stars with quadrupolar and dipolar internal fields.  

Considering, the complexities invloved in these simulations and
the high potential for these techniques to be used to understand
the dynamic behavior of neutron stars and other astrophysical systems,
it is prudent to develop different numerical schemes to cross verify
the results obtained. Moreover, having different reliable schemes gives
the flexibility to choose the one which can be most easily adapted for
a particular application. 
By far, the best advantage is of obtaining different physical
insight when the problem is solved for different settings.

In this work, we present and implement a numerical magneto--hydrodynamic
scheme to study the dynamics of differentially rotating proto--neutron
stars.  
Assuming the proto--neutron star to be composed of an
incompressible homogeneous fluid, we evolve the toroidal components
of the magnetic field and flow for fixed poloidal components.  The star is
assumed to be spherical in shape and the kinetic energy of rotation
is assumed to be far lesser than the magnetic energy which itself is far 
lesser than the gravitational energy.
The scheme is tested against numerical dissipation of physically
conserved quantities and verified by comparing the results obtained with
those from similar numerical computations (Liu \& Shapiro 2004). Although
the model used to represent the proto-neutron star is simplistic, we
show that qualitative insights on the dynamic behavior of such systems can 
be obtained. The motivation here is to present a reliable numerical scheme 
which may be further developed to simulate realistic neutron stars and other
astrophysical systems.

In the next section (\ref{sec:methodology}) the details of the technique are 
presented.  In section(\ref{sec:results}) the results of the simulations are 
given which are discussed in section (\ref{sec:discussion}). Appendix A
provides the details of numerical implementation of the equations.

\section{Methodology}
\label{sec:methodology}

We assume the differentially rotating star to be made up of a homogeneous,
incompressible, perfectly conducting fluid.  The relevant Magnetohydrodynamic 
(MHD) equations are:

\begin{eqnarray}
\mathbf{\nabla}\cdot\mathbf{v} & = & 0  \\
\frac{\partial {\mathbf v}}{\partial t}  & + &
({\mathbf v}\cdot{\mathbf \nabla}){\mathbf v}   =  
-\frac{1}{\epsilon} \mathbf{\nabla} P
-\mathbf{\nabla} \Phi
+\nu \nabla^{2} {\mathbf v}
+\frac{(\mathbf{\nabla} \times \mathbf{B})\times \mathbf{B}}{4 \pi \epsilon}
\nonumber \\
\\
\nabla^2 \Phi & = & 4 \pi G \epsilon     \\
\mathbf{\nabla}\cdot\mathbf{B} & = & 0 \\
\frac{\partial \mathbf{B}}{\partial t} & = &
\mathbf{\nabla} \times (\mathbf{v} \times \mathbf{B})
\end{eqnarray}
where $\vec{v}$ is the fluid velocity, $P$ the material pressure, 
$\epsilon$ the matter density, $\Phi$ the gravitational potential, 
$\nu$ the kinematic viscosity and $\vec{B}$ the magnetic field.

For our purpose here, we assume the star to be spherically symmetric 
(with a radius $R$) and immersed in an ambient plasma.  The initial 
differential rotation is 
taken to be constant along cylinders, and the poloidal component of the 
internal magnetic field to be time--independent and given to be a pure
dipole:
\begin{eqnarray}
B^{\rho}(t,\rho,z) & = & \frac{\mu}{R^3} \frac{3 \rho z}{(\rho^2 + z^2)^{5/2}} 
\\ \nonumber
B^{z}(t,\rho,z) &= &\frac{\mu}{R^3} \frac{(2 z^2 - \rho^2)}
{(\rho^2 + z^2)^{5/2}}
\end{eqnarray}
where $\mu$ is the magnetic moment. We make use of cylindrical coordinates: 
$\rho$, $z$ (scaled with respect to $R$) and $\varphi$ here
(the spherical symmetry of the star not withstanding, the 
quantities that are being considered in this problem possess axial 
symmetry in general).
Such a magnetic field being divergence and curl free, nonetheless, represents 
a field in vacuum (rather than one that resides inside a medium as is
considered here) and contains a singularity at the centre.  Since this field 
is time--independent, and we 
assume a purely axial flow, the central singularity in the magnetic field does 
not induce irregular behaviour in the velocity field (even though in the final 
state, the rotation rate will have a singularity at the centre) or the 
azimuthal magnetic field.  In the event of performing more realistic 
calculations, with more realistic magnetic field structures (such as the 
output of gravitational collapse codes), the numerical techniques developed 
here for treating such a restrictive magnetic field strucutre (the 
central singularity turns up as a stringent Courant condition) 
is expected to prove handy.

In order to compare with previous calculations in the literature, we use the 
same initial rotational and toroidal magnetic field profiles as that of 
Shapiro (2000), viz.:
\begin{eqnarray}                
\Omega(0,\rho,z) & = & \frac{1}{2} \Omega_0[1 + cos(\pi \rho^2)] \\
B^{\varphi}(0,\rho,z) &= &0 \\ 
\label{eq:omega0}
\end{eqnarray}                
where $\Omega_0$ represents a characteristic rotation rate at the
centre.

For our assumptions to be self consistent, we require 
$T \ll {\mathcal M} \ll W$,  where $T$ is the rotational kinetic 
energy, ${\mathcal M}$, the total magnetic energy and $W$ the 
gravitational potential energy.  In other words, for the magnetic
field structure that we assume here, at time $t>0$ 
\begin{eqnarray}
(B^{\rho})^2 + (B^{z})^2 \gg (B^{\varphi})^2 &  \sim & 
4 \pi \epsilon \rho^2 \Omega^2 \nonumber 
\end{eqnarray}
Since there is no mass in--flow to the system and since the system is 
homogeneous the poloidal equations decouple from the toroidal component 
of the Navier--Stokes equation (Shapiro \& Teukolsky 1983).  
We also assume that viscous dissipation, turbulence, convection and 
diffusion are absent in the system.  Another tacit assumption is
that the system is stable against the innumerable plasma/MHD 
instabilities (Spruit 1999).

\subsection{Flow and Magnetic Field Parameters}

\subsubsection{Formalism}
For our treatment here, we make use of the natural coordinates
available to us, viz., one that follows the magnetic field lines.
For the sake of comparison with the equivalent 1--D problem, we use 
a modified version of magnetic flux coordinates by defining 
the magnetic field as
\begin{eqnarray}
{\mathbf B} & = & {\mathbf \nabla} \psi \times {\mathbf \nabla} \varphi 
\end{eqnarray}
To be consistent with our assumptions of the time independent poloidal 
field dominating over the time dependent toroidal field, the equality above 
is only approximate.  With such a definition, we further choose
\begin{eqnarray}
\hat{\psi} & = & \psi/(\mu R)  =  \frac{\rho^2}{(\rho^2+z^2)^{3/2}} \\
\hat{\varphi} & = & \varphi  \\
\rho & = & \rho
\end{eqnarray}
to be the first, second and third coordinate respectively.  These choices 
are such that $\hat{\psi}$ is constant along a magnetic field line.
The inverse transformations are: 
\begin{eqnarray}
z & = & \sqrt{\left(\frac{\rho^2}{\hat{\psi}}\right)^{2/3} - \rho^2}
\\
\varphi & = & \hat{\varphi} \\
\rho & = & \rho
\end{eqnarray}

In the following paragraphs, we drop the `carets' from the variables
and tacitly understand the variables to be dimensionless.  With this, 
treatment, we have for the new coordinate system

\begin{eqnarray}
{\mathbf B} & = & (B^{\psi},0,B^{\rho})  \\
B^{\psi} & = & 0 \\
h_\psi & = & \sqrt{g_{\psi \psi}} = 1/(\rho B^{\rho}) \\
h_\varphi & = & \sqrt{g_{\varphi \varphi}} = \rho \\
h_{\rho} & = & \sqrt{g_{\rho \rho}} = B/B^{\rho} \\
g_{\psi \rho} & = &  B^z/(\rho (B^{\rho})^2) \\
g_{\psi \varphi} & = & 0 = g_{\rho \varphi} \\
g & = & g_{\varphi \varphi} (g_{\psi \psi} g_{\rho \rho} - g_{\psi \rho}^2) 
= 1/(B^{\rho})^2
\end{eqnarray}
\noindent we have again, as earlier, made the tacit assumption of 
neglecting the $B^{\varphi}$ in comparison to the poloidal
components.  In the above equations, 
$B= \sqrt{(B^{\rho})^2 + (B^z)^2}$ is the magnitude
of the magnetic field, $g_{i j}$ and $h_{i}$ (with 
$i$,$j = \psi$, $\varphi$, $\rho$) are the metric coefficients
and scale factors respectively and $g$ is the determinant of the metric
tensor.  On a single field line, the metric reduces to:
\begin{eqnarray}
d\hat{s} & =&  h_\rho d\rho
\label{eq:metric}
\end{eqnarray}
where $\hat{s}$ is the length along the magnetic field line.  Equation
(\ref{eq:metric}) agrees with the equation of a magnetic field
line (D'haeseleer et al. 1991). Integrating
this last equation gives the length of the magnetic field line 
$\hat{s}_{\rm max}(\psi)$.  
\begin{figure}
[htbp]
\begin{center}
\vspace{0.5cm}
\hspace{-0.5cm}
 \includegraphics[width=0.5\textwidth]{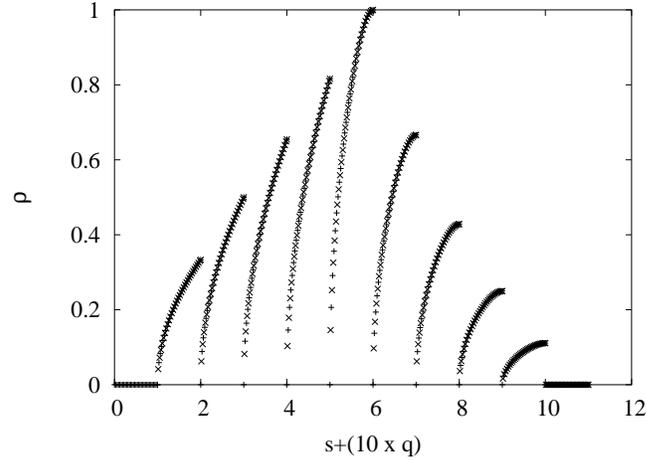}
\end{center}
\caption{The variation of $\rho$ with respect to
$s+(10 \times q)$ for a dipolar magnetic field.  This and other figures
(unless explicitly mentioned) employ 11 values of $q$; each value
corresponding to a field line.  For each line, the value of $q$ can be read 
off from the x-axis of this figure (since $0<s<1$ as 
$0<\rho<\rho_{\rm max}$, where
$\rho_{\rm max}$ corresponds to the value of the cylindrical radial coordinate at
the surface of the star).  
The lowest value of q corresponds to an `open' field line almost parallel
to the rotation axis while the largest one corresponds to a `closed' field
line enclosing the core of the star.  The line corresponding to $q =0.5$ is 
the marginally `closed' field line intersecting the equator at the surface
of the star.  Predictably, the non--linearity of the variation
increases with $q$. 
} 
\label{fig:srho}
\end{figure}

\begin{figure}
[hbp]
\begin{center}
\vspace{0.5cm}
\hspace{-0.5cm}
 \includegraphics[width=0.5\textwidth]{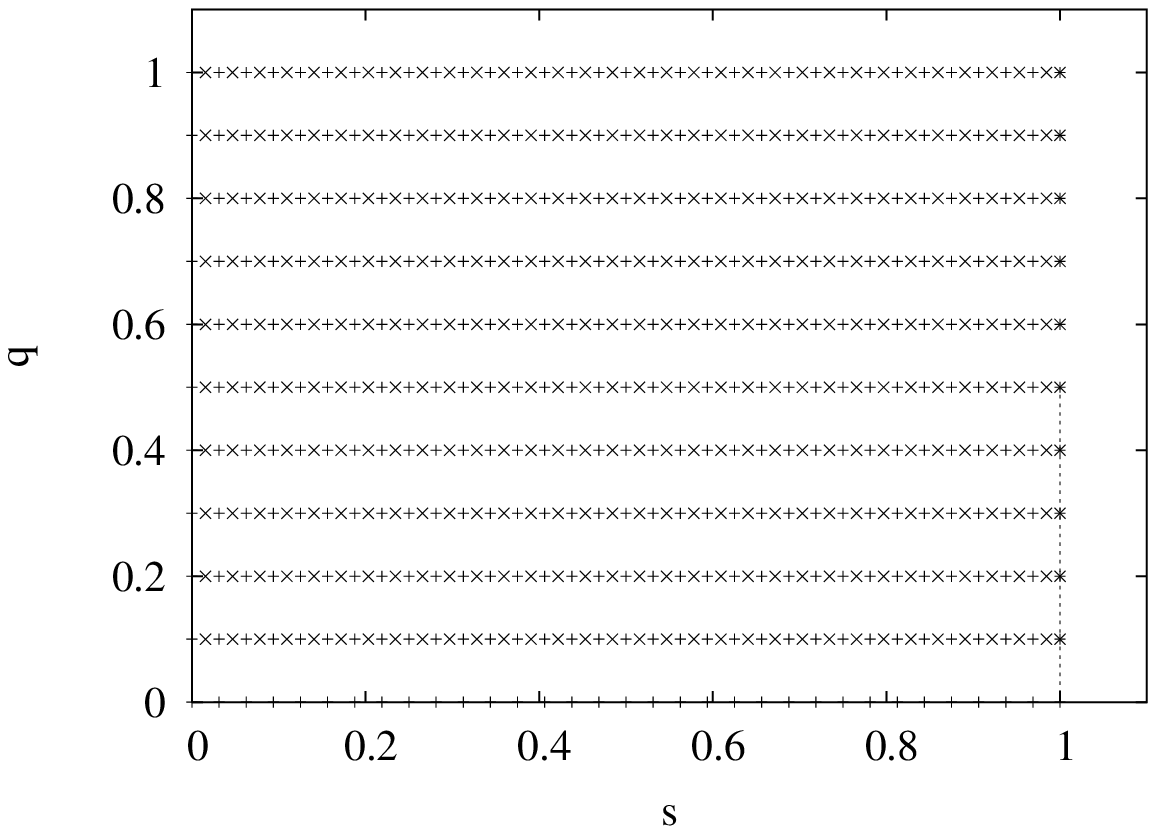}
 \includegraphics[width=0.5\textwidth]{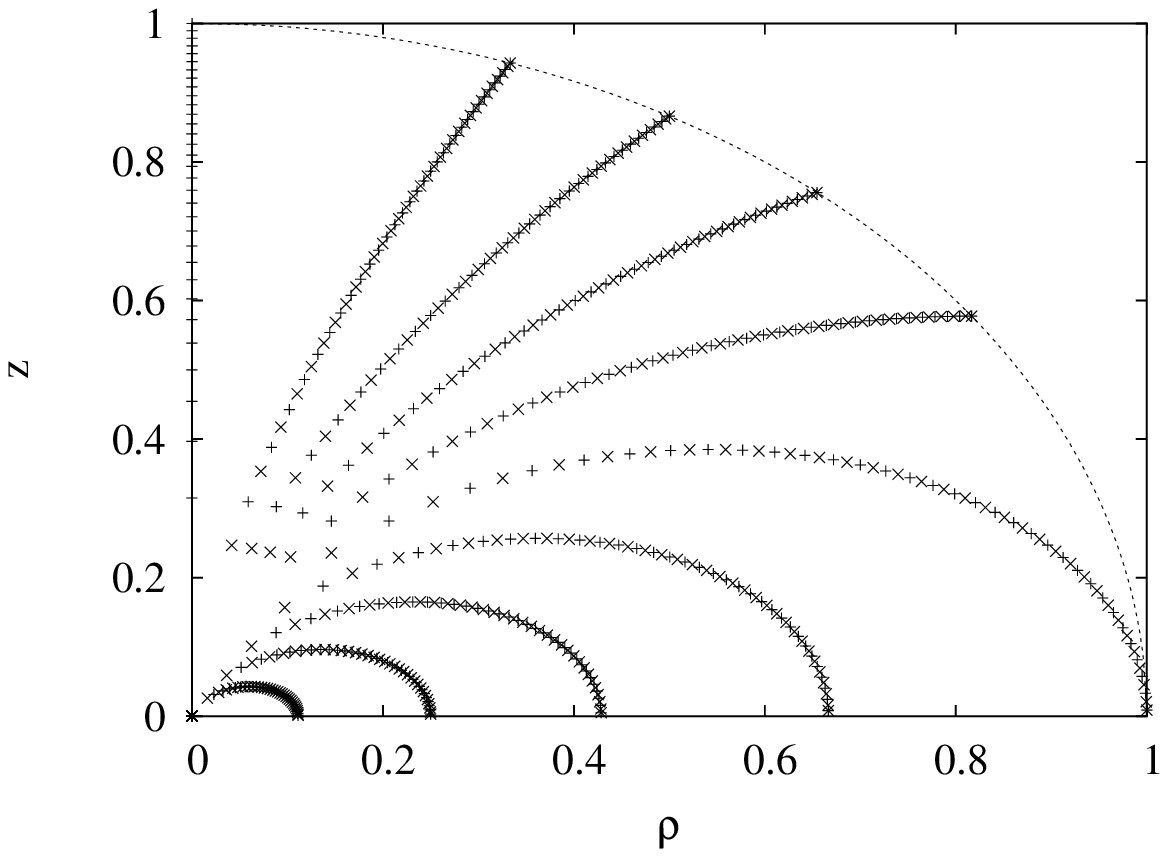}
\end{center}
\caption{Computational domain: in the left panel are the field lines
depicted by the `crosses' and the `pluses'
for $q$ v/s $s$; the vertical line represents the surface of the star
(the region $q>0.5$ represents the `closed' field line that never intersect
the stellar surface), the right panel displays the grid points projected 
onto the field lines in real space (cylindrical coordinates $z$ v/s
$\rho$) - the coordinate $q$ labels each of these lines. The `crosses'
correspond to `cell boundaries' at which $B^{\varphi}$ is defined,
while the `pluses' correspond to `cell centres' where $\Omega$
is defined in the staggered space scheme that we use here.}
\label{fig:comp_dom}
\end{figure}

In this new coordinate system, the azimuthal components of the 
Navier Stokes and magnetic induction equations reduce to:
\begin{eqnarray}
\frac{\partial v^{\varphi}}{\partial t} & = & 
\frac{1}{4 \pi \epsilon} \frac{h_{\psi} B^{\rho} h_{\rho}}{\sqrt{g} 
\hat{s}_{\rm max}(\psi)} 
\frac{\partial}{\partial s} (h_{\varphi} B^{\varphi}) 
\label{eq:mmfc1}\\
\frac{\partial B^{\varphi}}{\partial t} & = & 
\frac{h_{\varphi} h_{\rho}}{\sqrt{g} \hat{s}_{\rm max}(\psi)} 
\frac{\partial}{\partial s}(h_{\psi} v^{\varphi} B^{\rho}) 
\label{eq:mmfc2}
\end{eqnarray}
where we denote $s=\hat{s}/\hat{s}_{\rm max}(\psi)$.  
These are the 
counterparts of the 2--D equations (in cylindrical coordinates) given below:
\begin{eqnarray}
\frac{\partial v^{\varphi}}{\partial t} & = &
+\frac{1}{4 \pi \epsilon}  
\left[\frac{B^{\rho}}{R \rho} 
\frac{\partial}{\partial \rho} (\rho B^{\varphi}) + 
\frac{B^{z}}{R \rho} 
\frac{\partial}{\partial z} (\rho B^{\varphi})\right]
\label{eq:vphicyl}\\
\frac{\partial B^{\varphi}}{\partial t} & = & \frac{1}{R}
\left[
\frac{\partial}{\partial z} (v^{\varphi} B^{z})
+\frac{\partial}{\partial \rho} 
(v^{\varphi} B^{\rho}) \label{eq:bphicyl}
\right]
\end{eqnarray}

\subsubsection{Computational Domain}
The central singularity in the magnetic field now shifts to a coordinate
singularity for $\psi$.  Numerically, we take care of this singularity by 
introducing a new variable $q$ such that $\psi=q/(1-q)$.  We make another
numerically convenient change of variable by integrating Eq. (\ref{eq:metric}) 
to obtain the length of the magnetic field line ($\hat{s}_{\rm max}$) and 
then scaling the variable $\hat{s}$ with this 
quantity to obtain a new coordinate $s= \hat{s}/\hat{s}_{\rm max}$.
In Fig.~\ref{fig:srho} we display the result of such an integration and plot
 $\rho$ with $s+10 \times q$ for the dipolar magnetic field.
With these change of variables, the two independent spatial coordinates 
in the system of equations are $q$, and $s$ such that $0<q<1$ and 
$0<s<1$.  We have therefore, converted the curved field lines into straight 
field lines as shown in the left panel of Fig.~\ref{fig:comp_dom}.  
The projected field lines with the grid points are shown in 
Fig.~\ref{fig:comp_dom}, right panel.
Since the relationship between $z$ and $\psi$ is non--linear such that
$z$ blows up for low $\psi$ values, the central points for these values
of $\psi$ look sparse in Fig.~\ref{fig:comp_dom} right panel.
\begin{figure}
[htbp]
\begin{center}
\vspace{0.5cm}
\hspace{-0.5cm}
\includegraphics[width=0.4\textwidth]{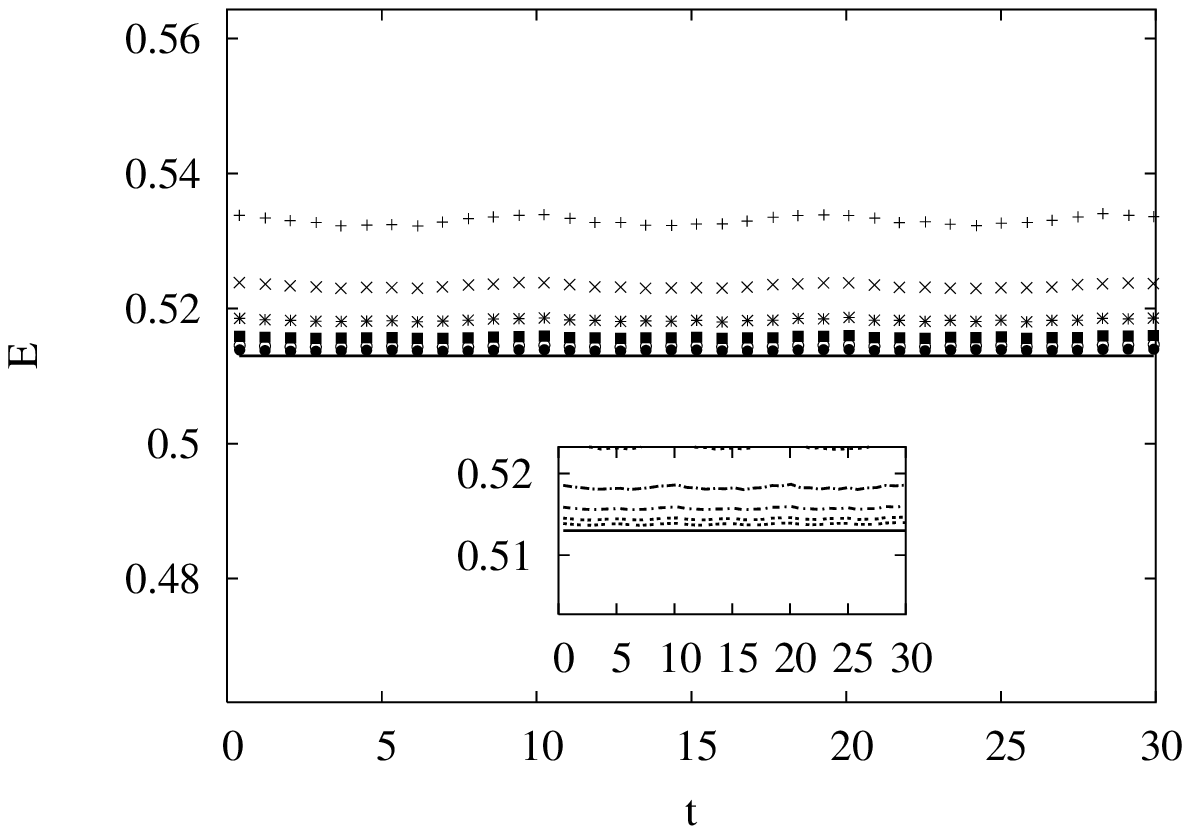}
\includegraphics[width=0.4\textwidth]{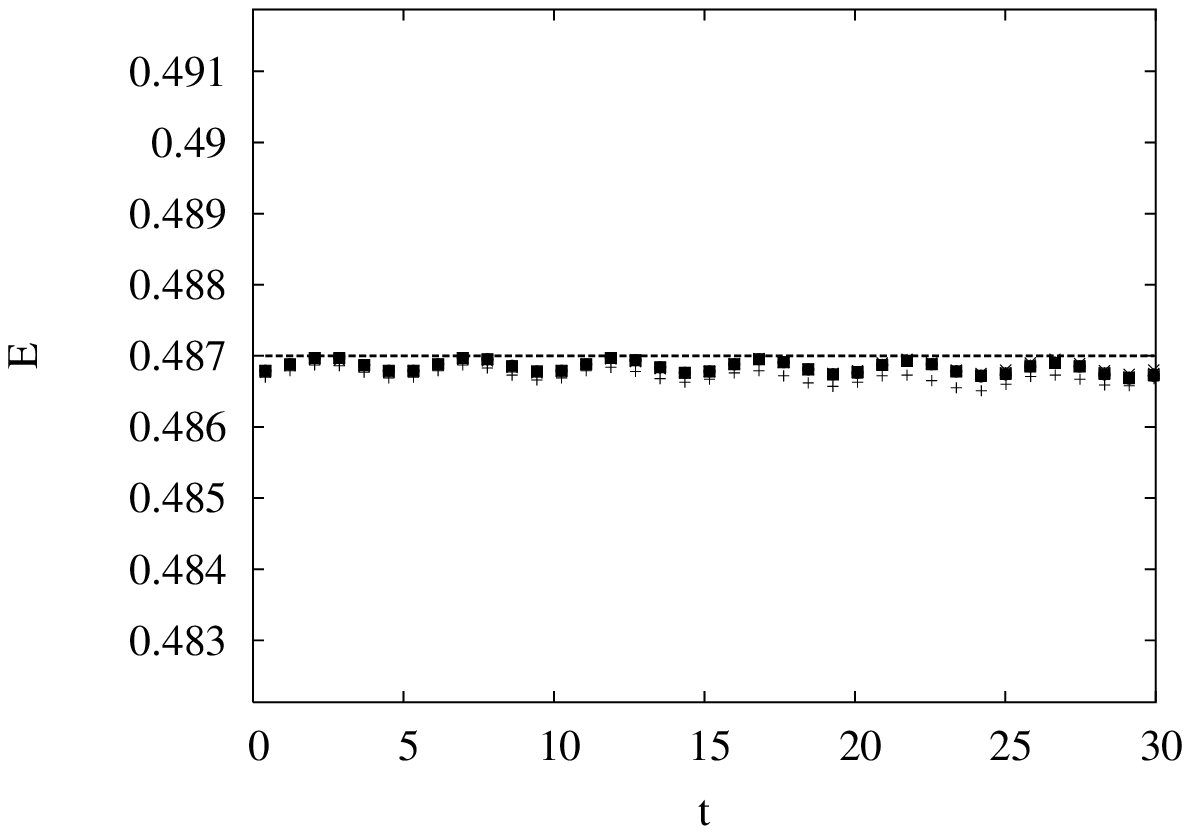}
\includegraphics[width=0.4\textwidth]{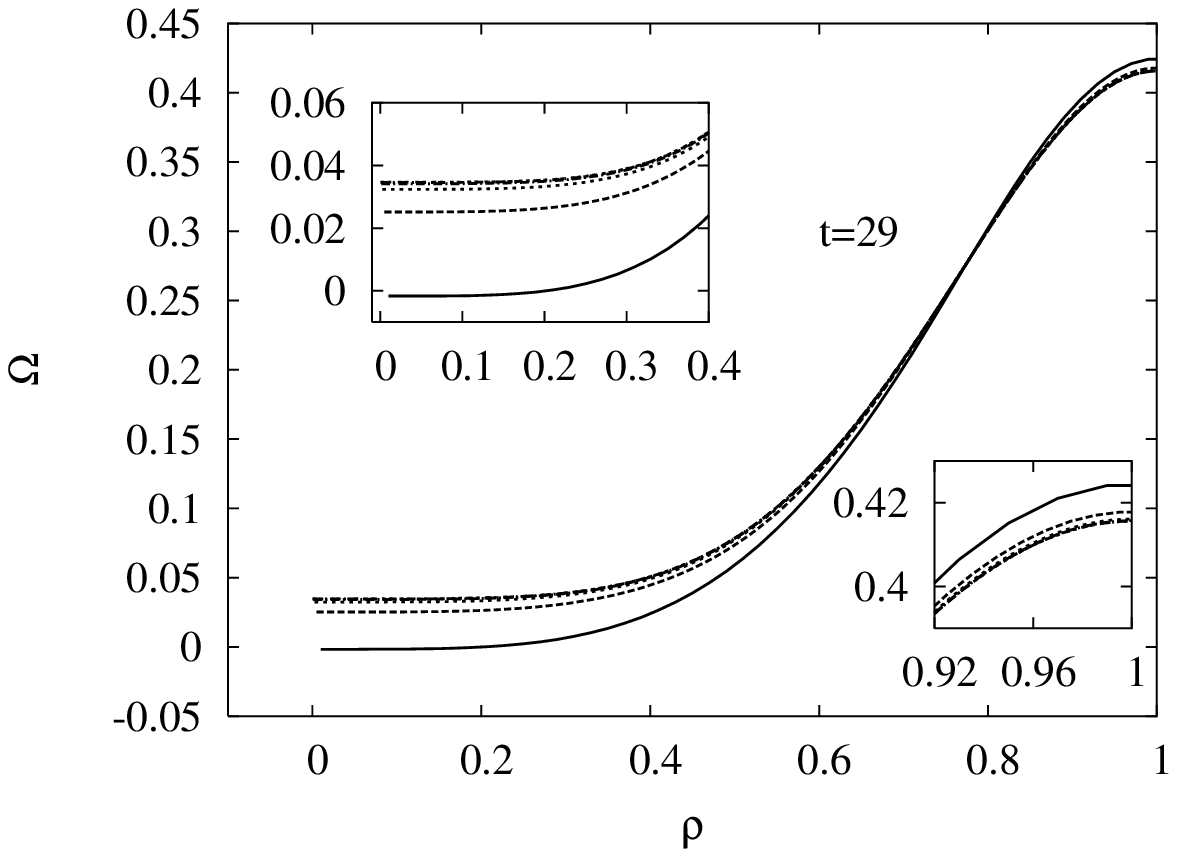}
\includegraphics[width=0.4\textwidth]{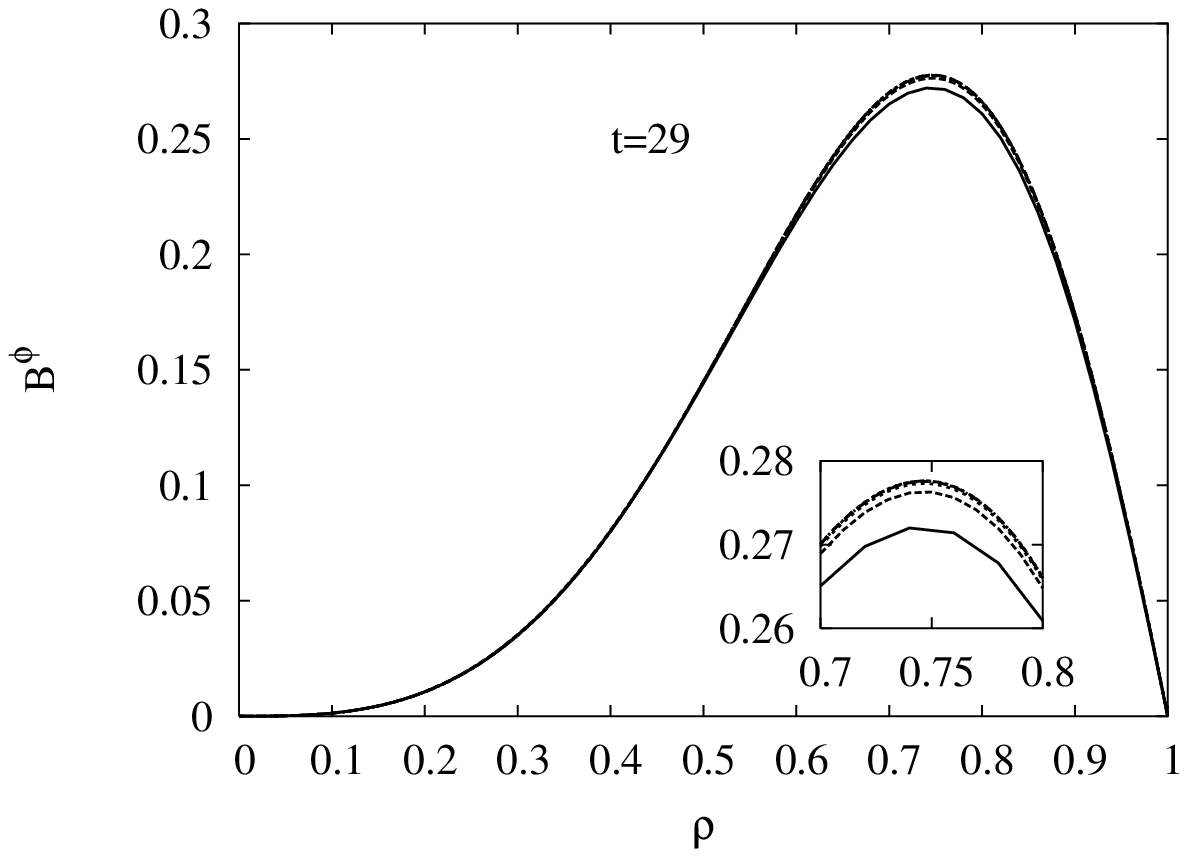}
\end{center}
\caption{
Convergence tests: For the monopolar field case, for vacuum exterior.
Top left panel is the convergence test for the minimum value of 
$E_{\rm rot}$: the straight line corresponding to $0.513$ is the analytically 
expected value; the limits on the y--axis represent $10$\% variation from
this value, the `+' are the numerically obtained values with 51 grid points, 
the `x' with 101 grid points, the '*' with 201, the filled squares with 401, 
the empty circles with 801 and the filled circles with 1601 (the last three 
appear merged in the main figure).  The inset shows a blown up view of the
same: solid line remains the same, the dashed line is with 101,
the long dash--dot line with 201, the medium dash--dot line with 401, 
the short dash--dot line with 801 and 
the dot--dot--dashed line with 1601 grid points respectively.
The top right panel is the maximum value of $E_{\rm mag}$, the 
analytically expected value of $0.487$ is represented by the straight
line; the limits of the y--axis correspond to a $1$\% variation.  
The numerically obtained values for these are fairly well coverged even
for $50$ grid point resolution.
The bottom left panel is the $\Omega$ profile at a late time of $=29$.
The curves have the same meaning as for the $E_{\rm rot}$ convergence.
The bottom right panel displays the $B^{\varphi}$ profile at $t=29$.
}
\label{fig:conv0}
\end{figure}

\subsubsection{Initial and Boundary Conditions}
\label{sec:bc}

We assume that the differential rotation is zero at the center.  This
amounts to saying that:
\begin{eqnarray}
\left.\frac{\partial \Omega}{\partial s} \right]_{(t,q,s=0)}& = & 0  \\
\left.\frac{\partial B^{\varphi}}{\partial t} \right]_{(t,q,s=0)}& = & 0 
\end{eqnarray}
For the outer boundary conditions, let us first consider the `closed' field 
lines.  From symmetry about the equator, we know that all the derivatives
in the vertical direction vanish in the equatorial plane, in addition
$B^{\rho} =0$ in the equatorial plane.  From 
Eq. (\ref{eq:bphicyl}) we see therefore that at $s=1$
\begin{eqnarray}
B^{\varphi} (t,q,s=1) & = & 0
\end{eqnarray}

For the `open' field lines, in a similar manner to Shapiro (2000), we
can show using Roberts (1968) that for our straight field lines
\begin{eqnarray}
\left.\frac{\partial B^{\varphi}}{\partial t} \right]_{(t,q,1)}& + & 
\left.\frac{\partial B^{\varphi}}{\partial s} \right]_{(t,q,1)} v_{\rm
A}
\frac{(1+{\cal R})}{(1-{\cal R})} = 0 \nonumber \\
\end{eqnarray}
where ${\cal R}$ is the reflection coefficient for the azimuthal magnetic
field amplitude:
\begin{eqnarray}
{\cal R}& = & \frac{\sqrt{(\epsilon_{\rm ex}/\epsilon)}-1}
{\sqrt{(\epsilon_{\rm ex}/\epsilon)}+1}
\end{eqnarray}
and $v_{\rm A} = B_0/\sqrt{4 \pi \epsilon}$ is the surface Alf\'{v}en
speed with $B_0=\mu/[h_{\varphi}]^3$ being the representative surface 
magnetic field.  It must be noted here that the actual internal Alf\'{v}en 
speed, $v^0_{\rm A} = B(\rho=0,z=0)/\sqrt{4 \pi \epsilon} >> v_{\rm A}$.

For the initial conditions, as mentioned earlier, we assume that  
$\Omega(0,q,s)$ is given by equation (\ref{eq:omega0}).  We also assume 
the toroidal magnetic fields to be absent to begin with i.e. 
$B^{\varphi}(0,q,s)=0$.

\subsubsection{Non--Dimensional Formulation}

For the sake of numerical convenience, we scale the dimensions out of the 
equations through the following identifications:
\begin{eqnarray}
& & h_{\varphi} =  [h_{\varphi}]\hat{h}_{\varphi},
~~~h_{\psi}  =  [h_{\psi}]\hat{h}_{\psi},
~~~h_{\rho}  =  [h_{\rho}]\hat{h}_{\rho} ,
~~~g  =  [g] \hat{g},
~~~v^{\varphi}  =  [h_{\varphi}] \Omega, \nonumber \\  
& & \Omega  =  \Omega_0 \hat{\Omega},
~~~B^{\rho}  =  B_0 \hat{B}^{\rho},
~~~B^{\varphi}  =  \sqrt{4 \pi \epsilon} \Omega_0 [h_{\varphi}]
 \hat{B}^{\varphi},
~~~t  = 2 \hat{t}/([h_{\varphi}]/v_{\rm A}),
~~~\frac{[h_{\varphi}] [h_{\psi}] [h_{\rho}]}{\sqrt{[g]}} = 1, \nonumber \\
& & E = \left( \frac{4 \pi}{3} \epsilon
\frac{\sqrt{[g]}[\psi][h_{\varphi}]}{[h_{\rho}]} \right) 
\frac{[h_{\varphi}]^2 \Omega_0^2}{2}
\hat{E}, \nonumber \\
& & J = \left( \frac{4 \pi}{3} \epsilon
\frac{\sqrt{[g]}[\psi][h_{\varphi}]}{[h_{\rho}]} \right) 
[h_{\varphi}] \rho^2 \Omega_0
\hat{J} \nonumber  
\end{eqnarray}

\begin{figure}
[hbp]
\begin{center}
\vspace{0.5cm}
\hspace{-0.5cm}
\includegraphics[width=0.4\textwidth]{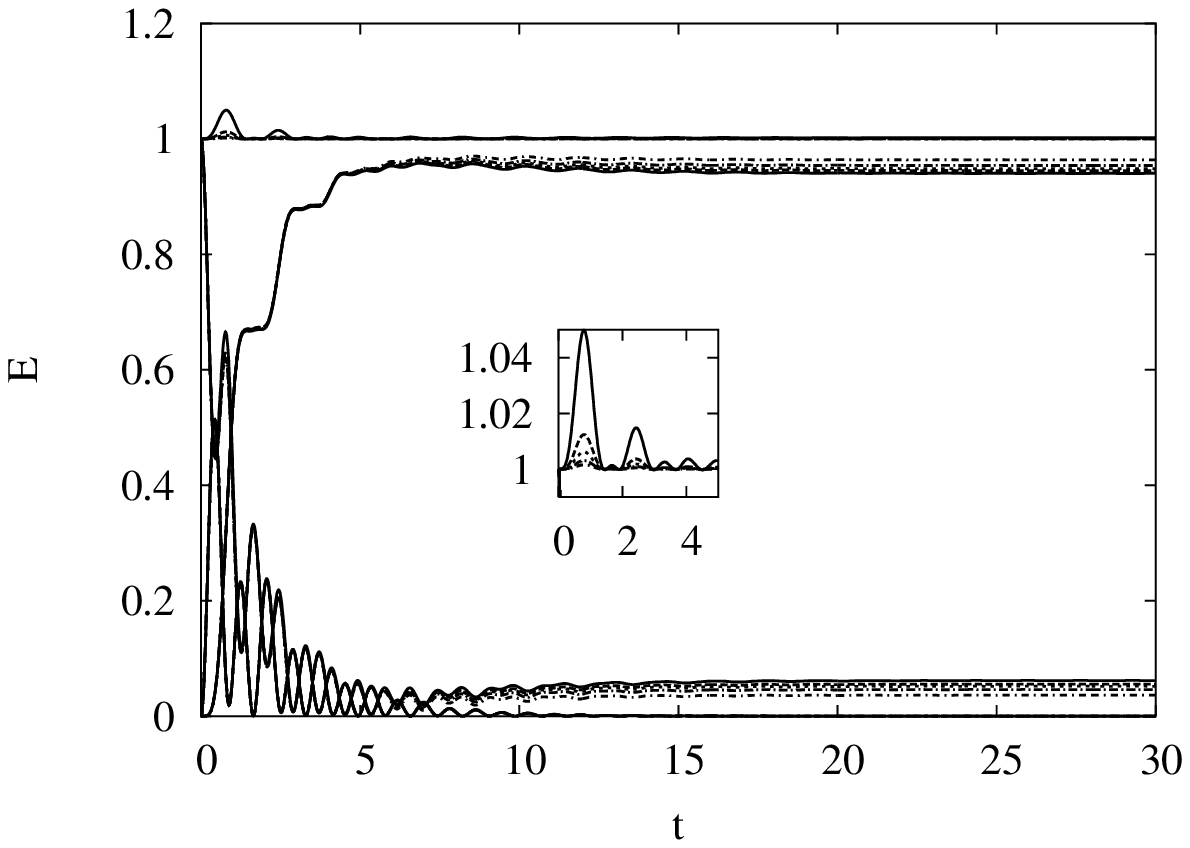}
\includegraphics[width=0.4\textwidth]{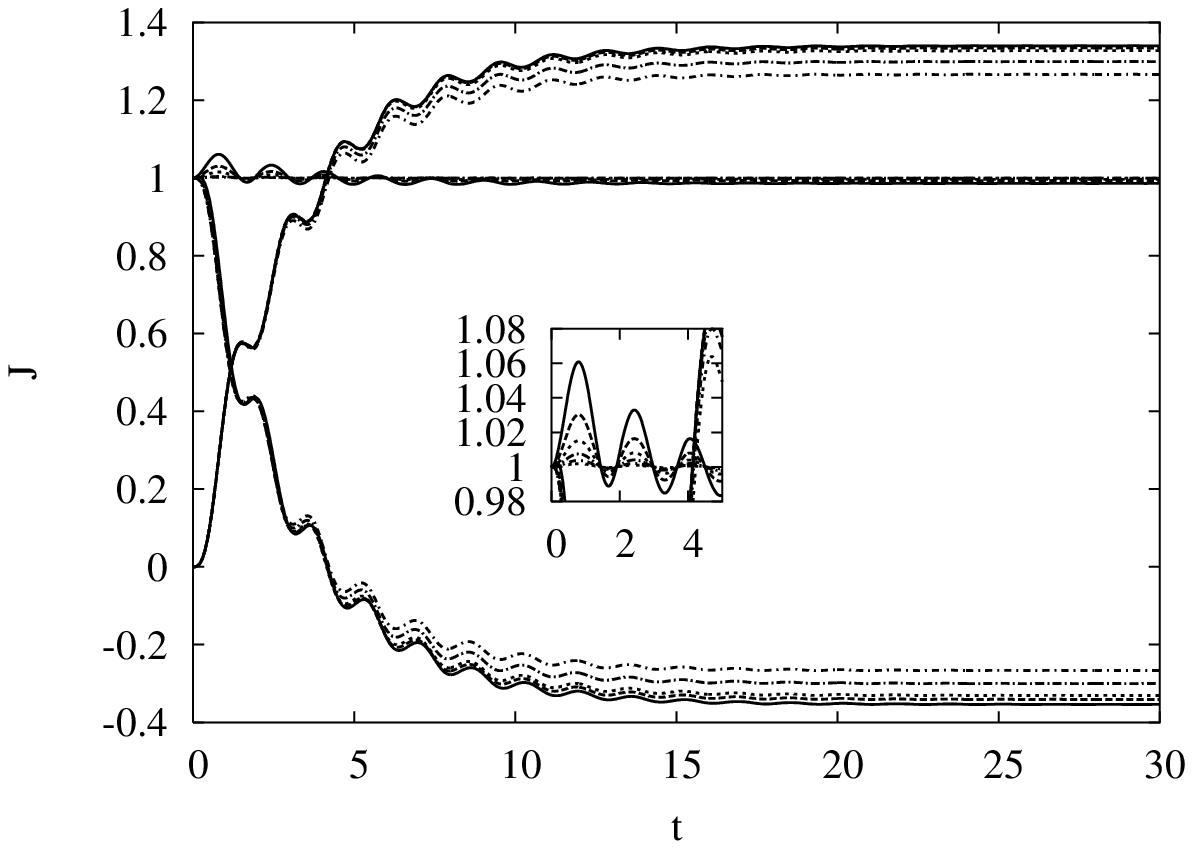}
\includegraphics[width=0.4\textwidth]{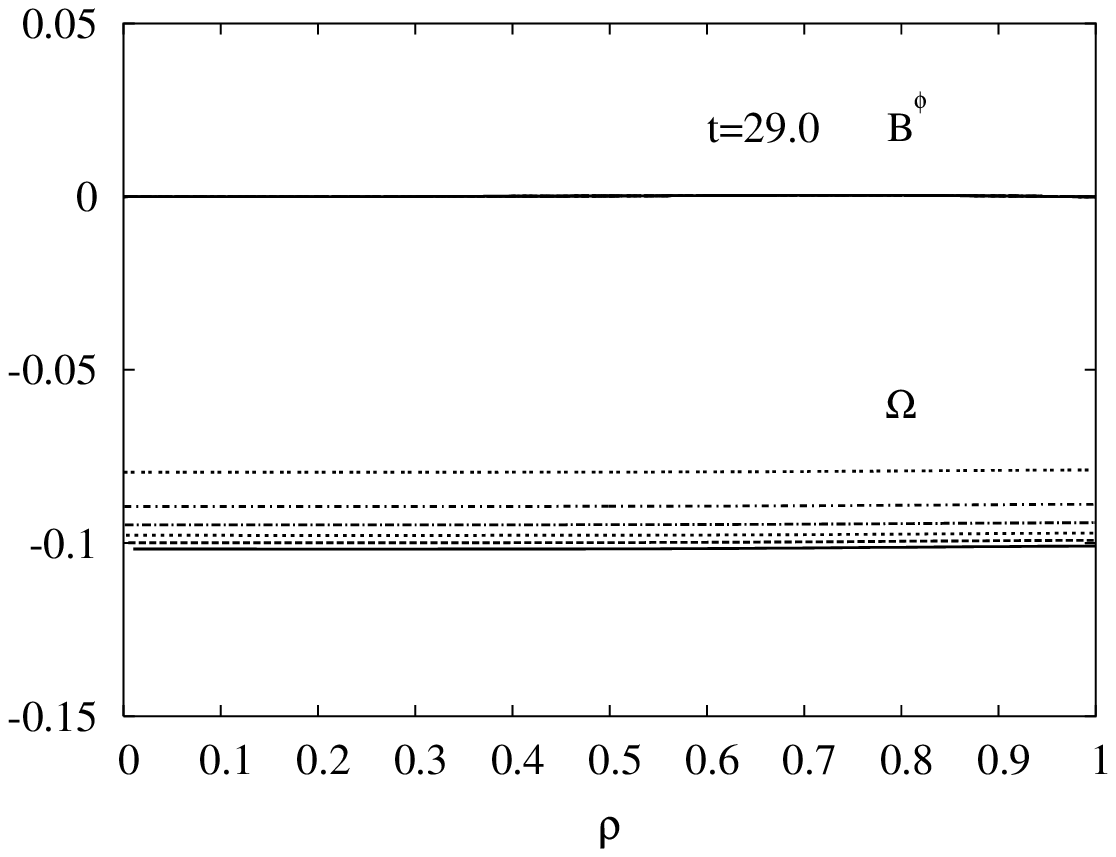}
\end{center}
\caption{
Convergence tests: For the monopolar field case, plasma exterior.
In the top left panel is displayed the energy conservation, the right one
displays angular momentum conservation while the bottom panel contains
the $\Omega$ and $B^{\varphi}$ profiles at time $t=29$.  In all panels, 
the solid, the dashed, the short--dashed, the dot--dashed, the
dot--short--dashed and the dotted lines indicate  resolutions of 
51, 101, 201, 401, 801 and 1601 points respectively. 
}
\label{fig:conv2}
\end{figure}

\subsubsection{Evolution Equations}
The non--dimensional form of the evolution Eqs. (\ref{eq:mmfc1}) and
(\ref{eq:mmfc2}) become:
\begin{eqnarray}
\frac{\partial \hat{\Omega}}{\partial \hat{t}} & = & 
\frac{\hat{h}_{\psi} \hat{B}^{\rho} \hat{h}_{\rho}}{2 \hat{h}_{\varphi}
\sqrt{\hat{g}} s_{\rm max}(q)}
\frac{\partial}{\partial s} 
\left(\hat{h}_{\varphi} \hat{B}^{\varphi}\right) 
\label{eq:evoln1}\\
\frac{\partial \hat{B}^{\varphi}}{\partial \hat{t}} & = & 
\frac{\hat{h}_{\varphi} \hat{h}_{\rho}}{2 \sqrt{\hat{g}} s_{\rm max}(q)}
\frac{\partial}{\partial s}
\left(\hat{h}_{\psi} \hat{h}_{\varphi} \hat{\Omega} \hat{B}^{\rho}\right)
\label{eq:evoln2}
\end{eqnarray}
where we have identified $v^{\varphi}$ with $h_{\varphi} \Omega$.

\begin{figure}
[htp]
\begin{center}
\vspace{0.5cm}
\hspace{-0.5cm}
\includegraphics[width=0.5\textwidth]{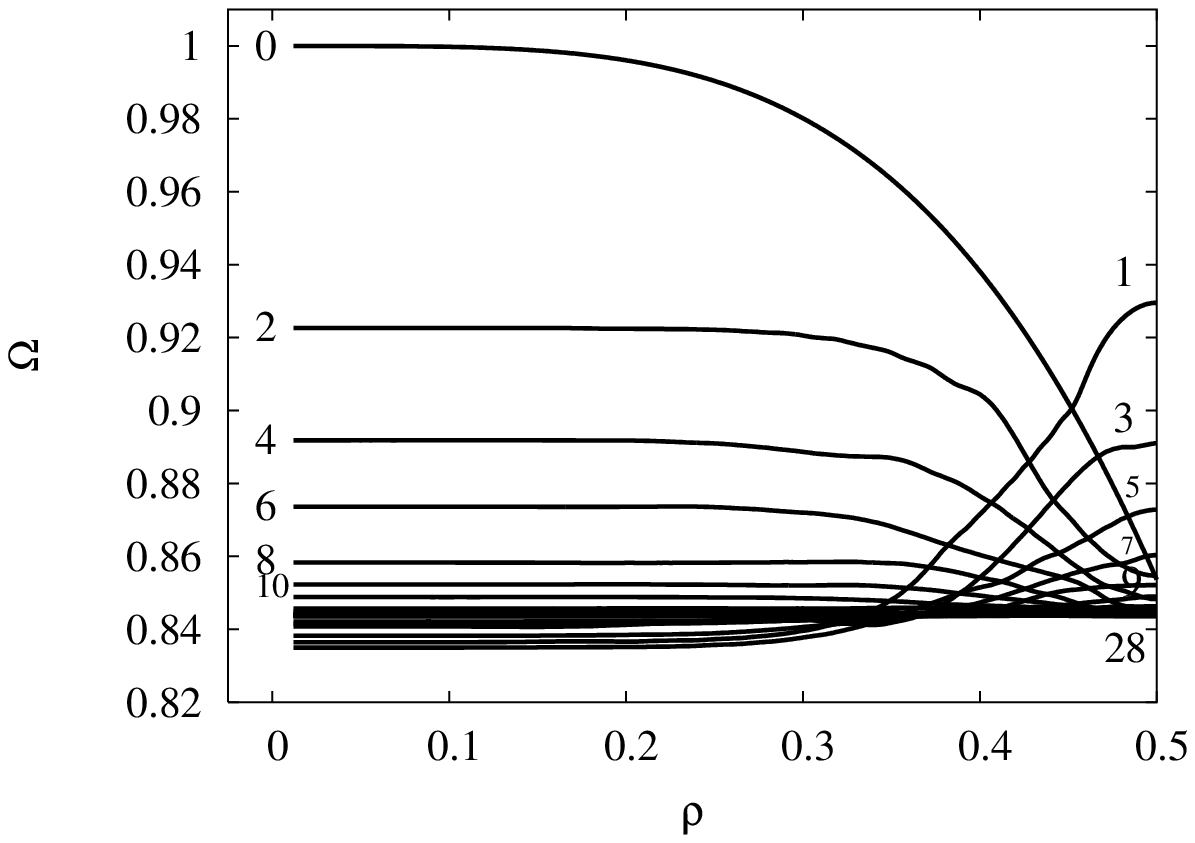}
\includegraphics[width=0.5\textwidth]{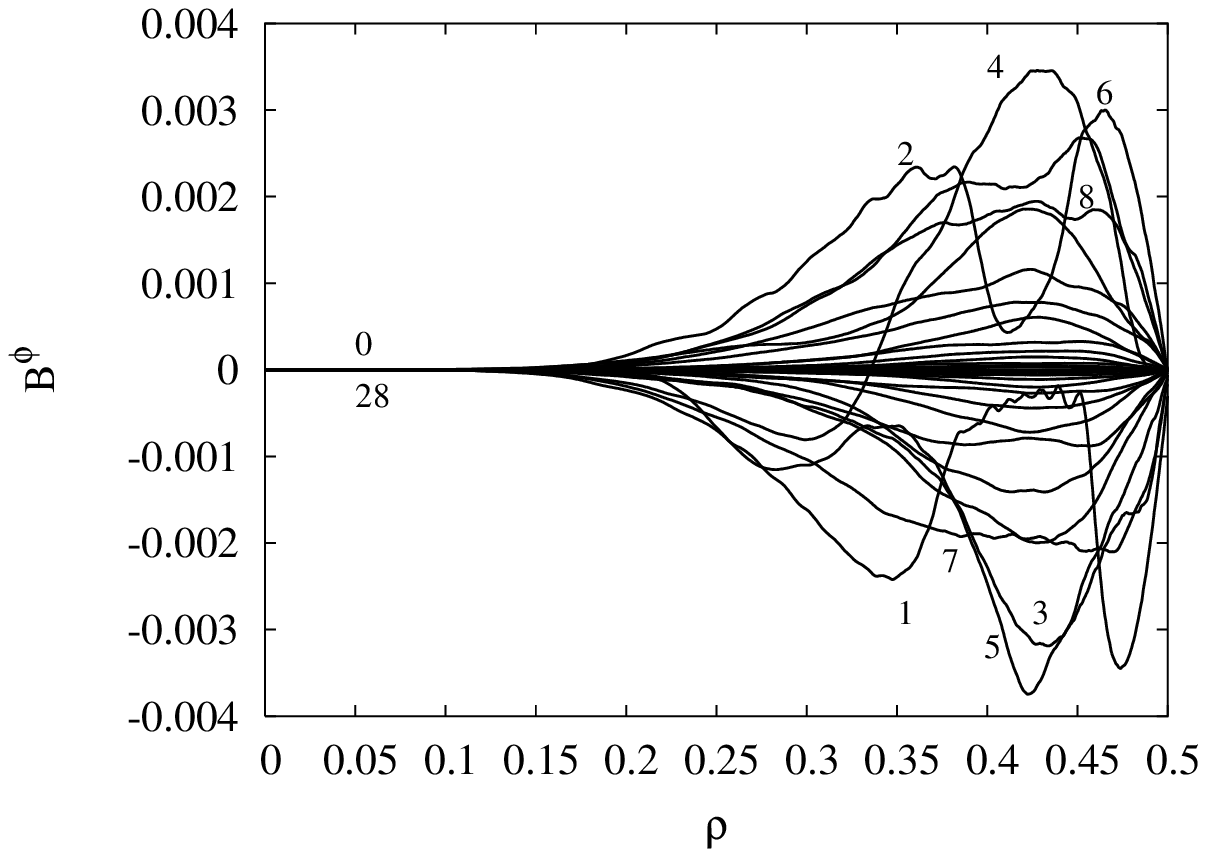}
\includegraphics[width=0.5\textwidth]{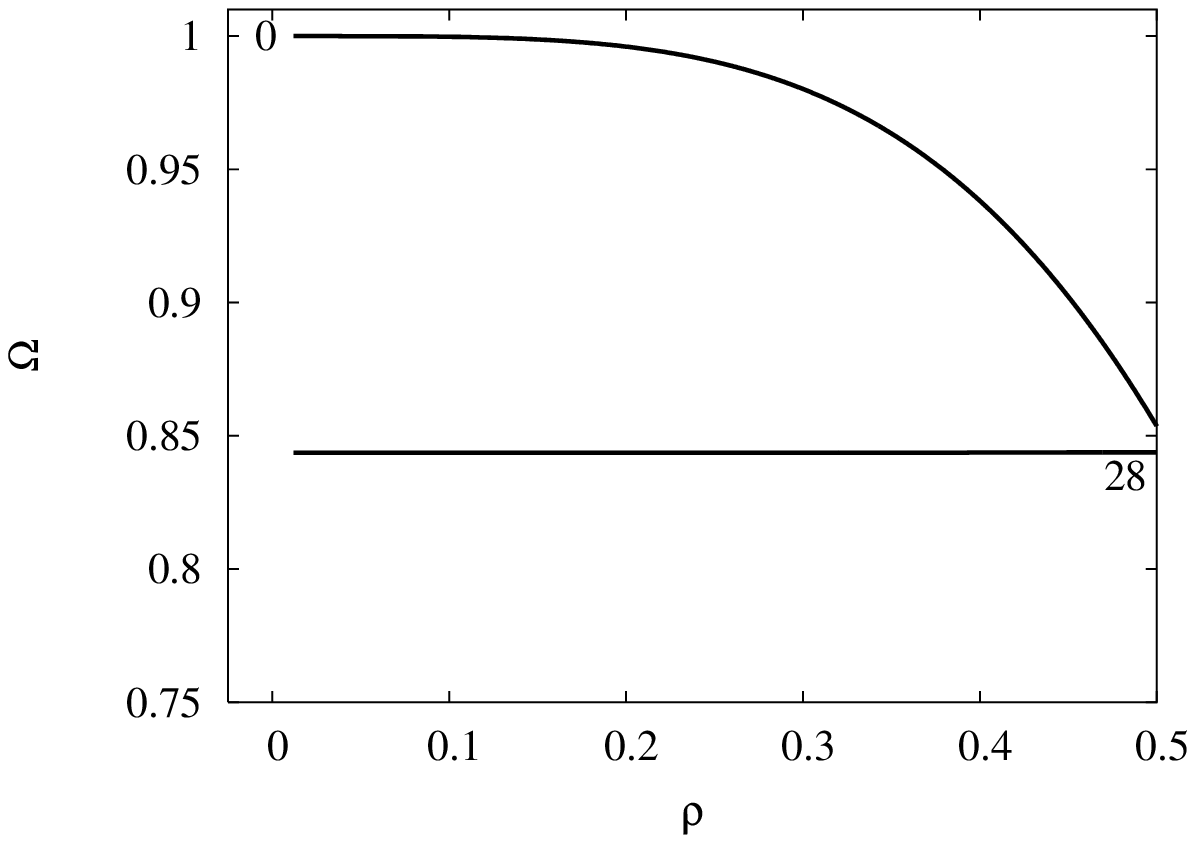}
\end{center}
\caption{$\Omega$ and $B^{\varphi}$ variation for the $q=0.3$ case. The first
panel shows the variation of $\Omega$ as a function of the cylindrical 
coordinate $\rho$ with time (the topmost curve being the initial curve 
and the rest the subsequent evolution).  The second panel displays 
$B^{\varphi}$ as a function of $\rho$ with time (the initial and final
profiles is a null profile and is hidden by the subsequent and earlier
evolution respectively). The third panel shows the initial and final
profile of $\Omega$ with $\rho$. The numbers labelling the curves
in this and other figures stand for the corresponding time $t$; in this
figure, since the curves get very close together at late times, we 
suppress labelling between $t=8$ and $t=28$.}
\label{fig:q3}
\end{figure}

\subsubsection{Energy and Angular Momentum Conservation}
As energy conservation acts as a check to the numerical computations, we
compute the relevant energies in the system.
In the non--dimensional units, the integrals for rotational kinetic energy, 
toroidal magnetic energy and Poynting energy are given as (in the equations
below, $d\tau$ represents the volume element in cylindrical coordinates):
\begin{eqnarray}
{\cal E}_{\rm rot}(t) & = & \int d \tau (\epsilon \Omega(t,\rho,z)^2 \rho^2/2)
= \int d \hat{\psi} \hat{E}_{\rm rot}
\\
{\cal E}_{\rm mag}(t) & = & \int d \tau [(B^{\varphi}(t,\rho,z))^2/8 \pi]
= \int d \hat{\psi} \hat{E}_{\rm mag} \\
{\cal E}_{\rm Poy}(t) & = & \int_0^t dt \dot{\cal E}_{\rm Poy}(t) 
= - \int_0^t dt \left( \int h_{\psi} h_{\varphi} d \psi d \varphi
\frac{B^{\varphi} B^{\rho} \Omega h_{\varphi}}{4 \pi}
\right)_{\hat{s}=\hat{s}_{\rm max}} 
=- \int_0^t dt 
\left(
\int d \psi \hat{\dot{\cal E}}_{\rm Poy}
\right)_{\hat{s}=\hat{s}_{\rm max}} \nonumber \\
& & ~~~~~~~~~~~~~~~~~~~~~~~~= \int d \hat{\psi} \hat{E}_{\rm Poy} 
\end{eqnarray}
where 
$\hat{\dot{\cal E}}_{\rm Poy}  = -\frac{1}{2} h_{\varphi} B^{\varphi}
\Omega$ and 
\begin{eqnarray}
\hat{E}_{\rm rot} & = & \frac{3}{2} \int ds \sqrt{\hat{g}} 
\frac{\hat{s}_{\rm max}}{\hat{h}_{\rho}} \hat{h}_{\varphi}^2 \hat{\Omega}^2
\label{eq:erot}
\\
\hat{E}_{\rm mag} & = & \frac{3}{2} \int ds \sqrt{\hat{g}} 
\frac{\hat{s}_{\rm max}}{\hat{h}_{\rho}} (\hat{B}^{\varphi})^2
\\
\hat{E}_{\rm Poy} & = & -\frac{3}{2} \int_0^{\hat{t}} d\hat{t} 
 \left( \hat{B}^{\varphi} \hat{\Omega} \hat{h}_{\varphi}
 \right)_{\hat{s}=\hat{s}_{\rm max}}
\end{eqnarray}
are the energies associated with a given magnetic field line.  Conservation
of energy requires

\begin{figure}
[htp]
\begin{center}
\vspace{0.5cm}
\hspace{-0.5cm}
\includegraphics[width=0.5\textwidth]{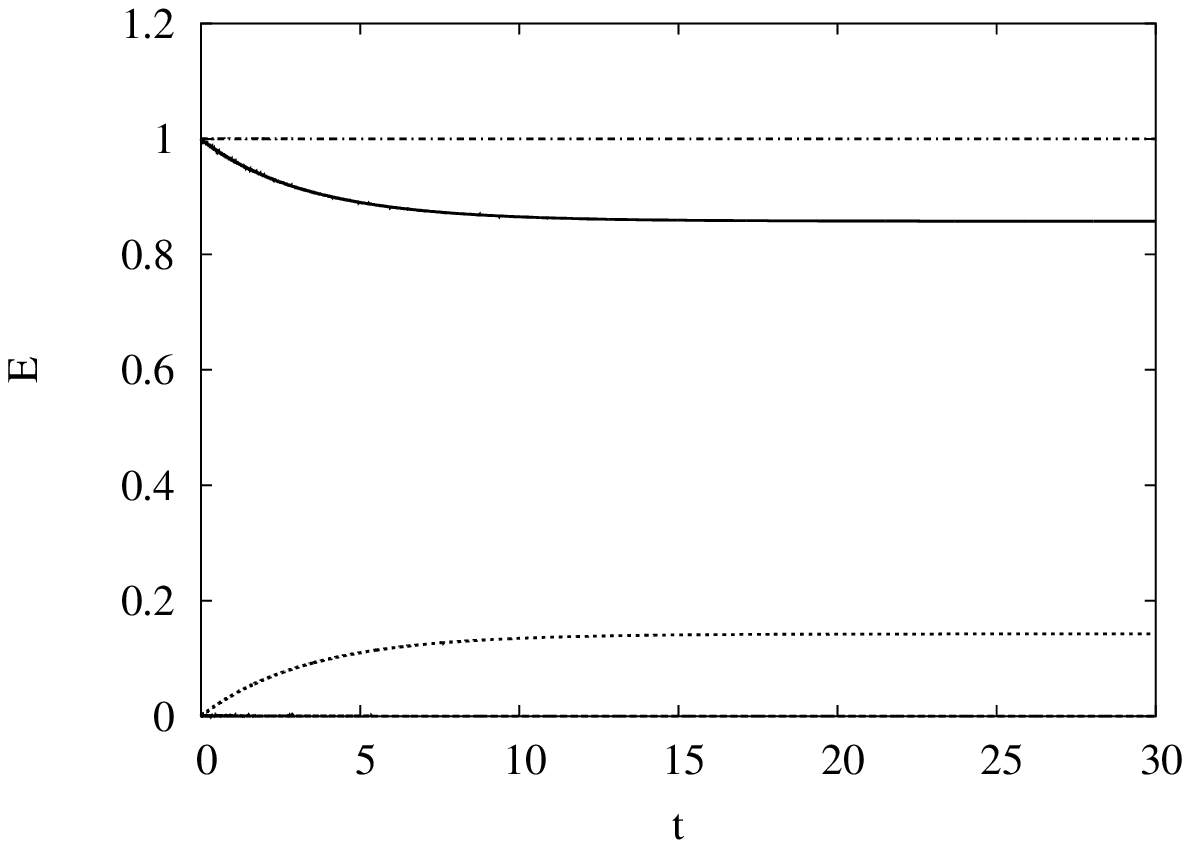}
\includegraphics[width=0.5\textwidth]{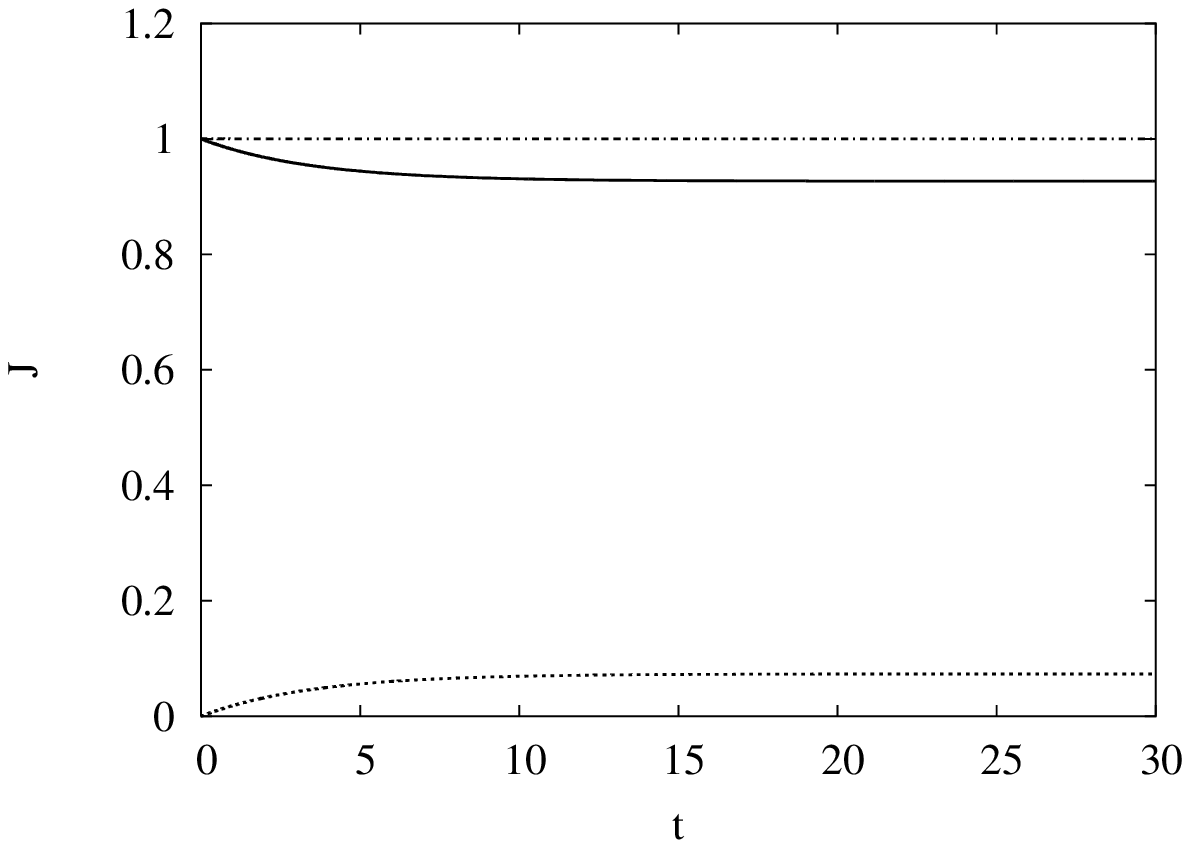}
\end{center}
\caption{Energy and angular momentum conservation for the $q=0.3$ case.
The left panel displays the energy conservation: the solid line is 
$E_{\rm rot}$, the long--dashed line (coincident with the x--axis) 
represents $E_{\rm mag}$, the short--dashed line is $E_{\rm Poy}$ and
the dot--dashed line is the sum of all these quantities (all energies
are normalised to the initial value of $E_{\rm rot}$).
The right panel
displays the angular momentum conservation: solid curve is $J_{\rm rot}$
(see text) while the short--dashed curve is the $J_{\rm Poy}$ and
the dot--dashed line is the sum (all scaled to the initial $J_{\rm rot}$).}
\label{fig:q3e}
\end{figure}

\begin{eqnarray}
{\cal E}_{\rm rot}(0) & =  & {\cal E}_{\rm rot}(t) + {\cal E}_{\rm mag}(t) +
{\cal E}_{\rm Poy} \label{eq:encons3d}
\end{eqnarray}

However, it may be noted that Eq. (\ref{eq:encons3d}) is satisfied
if on each magnetic field line:
\begin{eqnarray}
\hat{E}_{\rm rot}(0) & = & \hat{E}_{\rm rot}(t) + \hat{E}_{\rm mag}(t) +
\hat{E}_{\rm Poy}(t)
\end{eqnarray}
holds.

The angular momenta integrals are:
\begin{eqnarray}
{\cal J}_{\rm rot}(t) & = & \int d \tau (\epsilon \Omega(t,\rho,z) \rho^2)
= \int d \hat{\psi} \hat{J}_{\rm rot}
\\
{\cal J}_{\rm mag}(t) & = & \int_0^t dt {\cal N}
= \int d \hat{\psi} \hat{J}_{\rm mag} 
\end{eqnarray}
where 
${\cal N} = \hat{\dot{\cal E}}_{\rm Poy}/\Omega$ is the torque exerted
by the Maxwell stress at the surface and 
\begin{eqnarray}
\hat{J}_{\rm rot} & = & \frac{3}{2} \int ds \sqrt{\hat{g}} 
\frac{\hat{s}_{\rm max}}{\hat{h}_{\rho}} \hat{h}_{\varphi}^2 \hat{\Omega}
\\
\hat{J}_{\rm mag} & = & -\frac{3}{4} \int_0^{\hat{t}} d\hat{t} 
 \left( \hat{B}^{\varphi} \hat{h}_{\varphi}
 \right)_{\hat{s}=\hat{s}_{\rm max}}
\end{eqnarray}

\begin{figure}
[htp]
\begin{center}
\vspace{0.5cm}
\hspace{-0.5cm}
\includegraphics[width=0.5\textwidth]{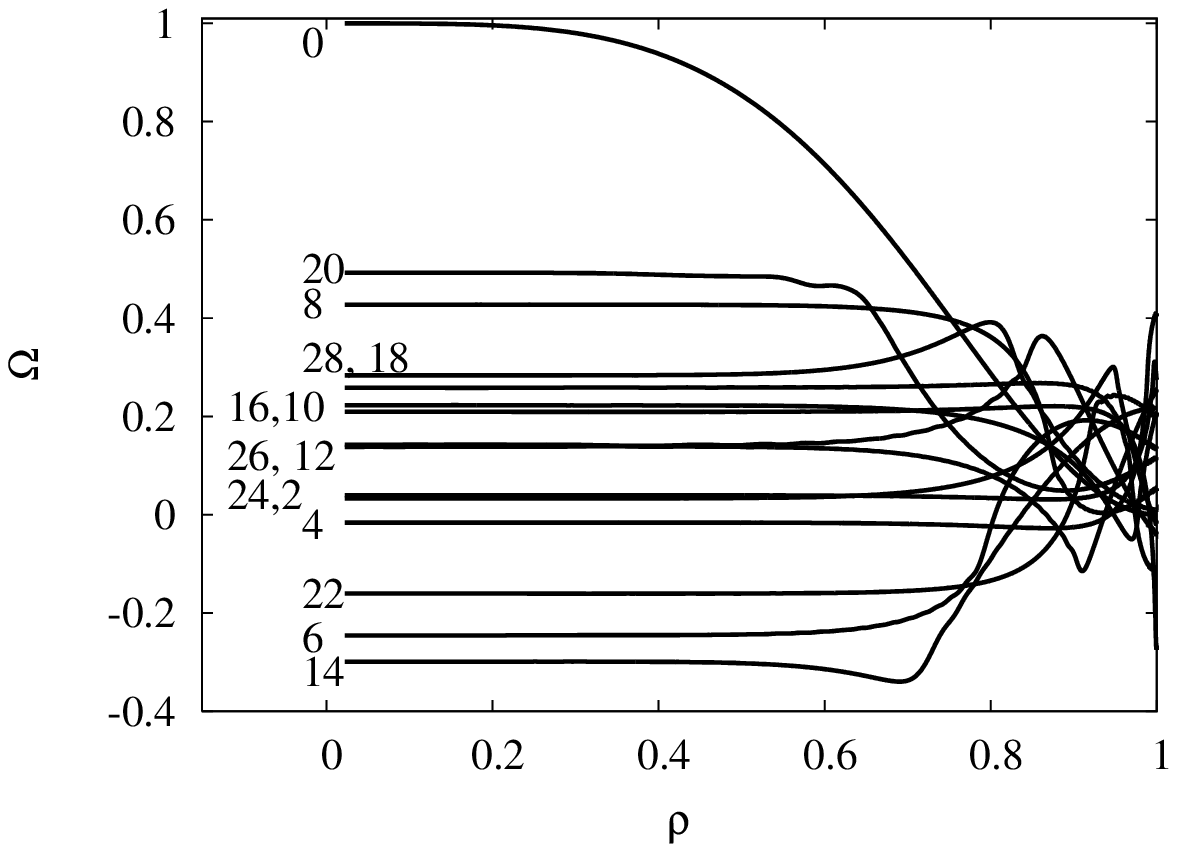}
\includegraphics[width=0.5\textwidth]{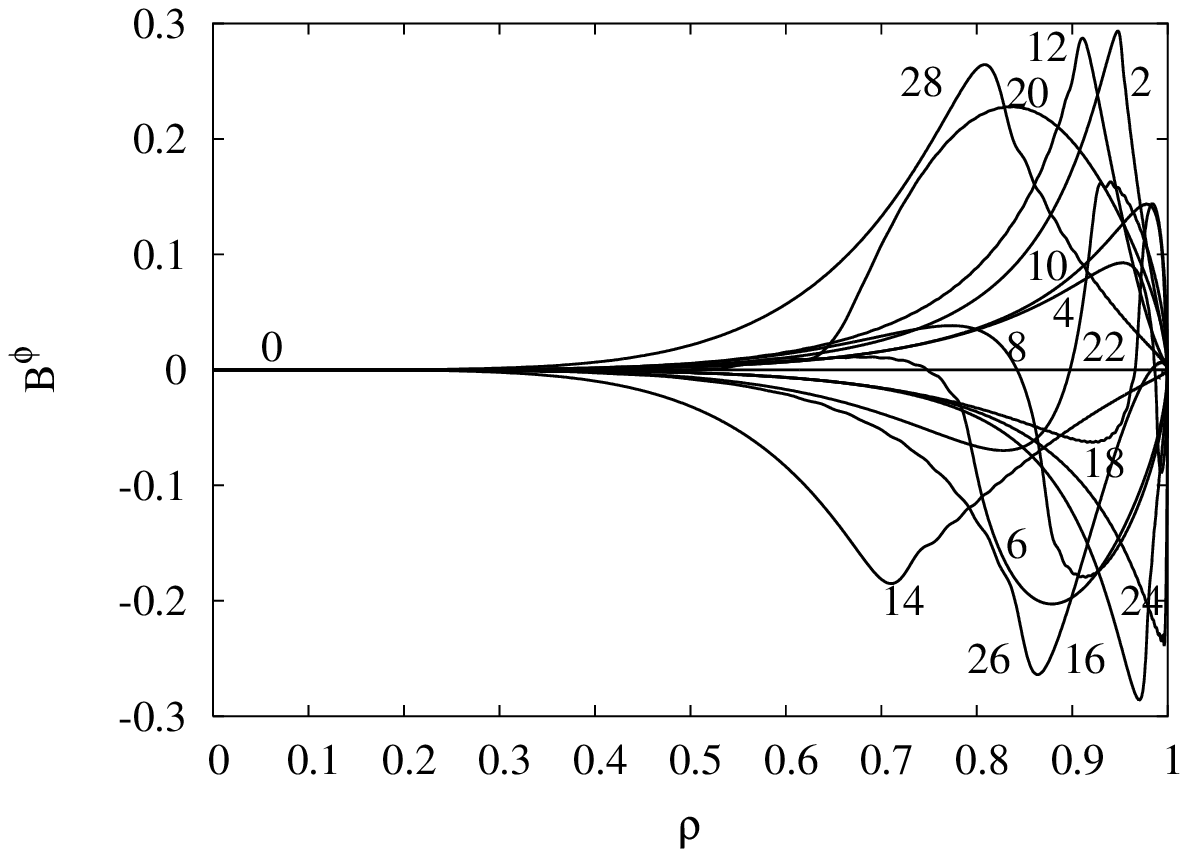}
\end{center}
\caption{$\Omega$ and $B^{\varphi}$ variation for the $q=0.6$ case.
The curves have the same meaning as for Fig.~\ref{fig:q3} except that
because $q=0.6$ is a closed field line, the oscillations are sustained.}
\label{fig:q6}
\end{figure}

\subsection{Numerical Scheme}

Since the coefficients to the differentials in Eqs. (\ref{eq:evoln1})
and (\ref{eq:evoln2}) depend on some powers ($>1$) of inverse of the 
coordinate distance from the centre, clearly, a fully explicit scheme 
may prove to be highly restritive due to the Courant condition.  For 
the equivalent 1--D problem, it was shown (Shapiro 2000) that for a 
fully implicit scheme employing a 2nd order staggered space and 1st order 
time, the Courant condition only proves to be a restriction for accuracy 
(numerical stability being assured) -- and the energy conservation (and 
angular momentum conservation in the case of energy dissipation due to 
an external plasma) is the marker for the accuracy.  Accordingly, for
our purpose here we use a staggered leapfrog scheme.
The recipe for the finite differencing for such a scheme is 
mentioned in the Appendix.

The details on the numerical checks performed on the code 
{\it vis--\'{a}--vis} 
the one--dimensional formalism of Shapiro (2000) is provided 
in section (\ref{sec:results}) below.

\begin{figure}
[htp]
\begin{center}
\vspace{0.5cm}
\hspace{-0.5cm}
\includegraphics[width=0.5\textwidth]{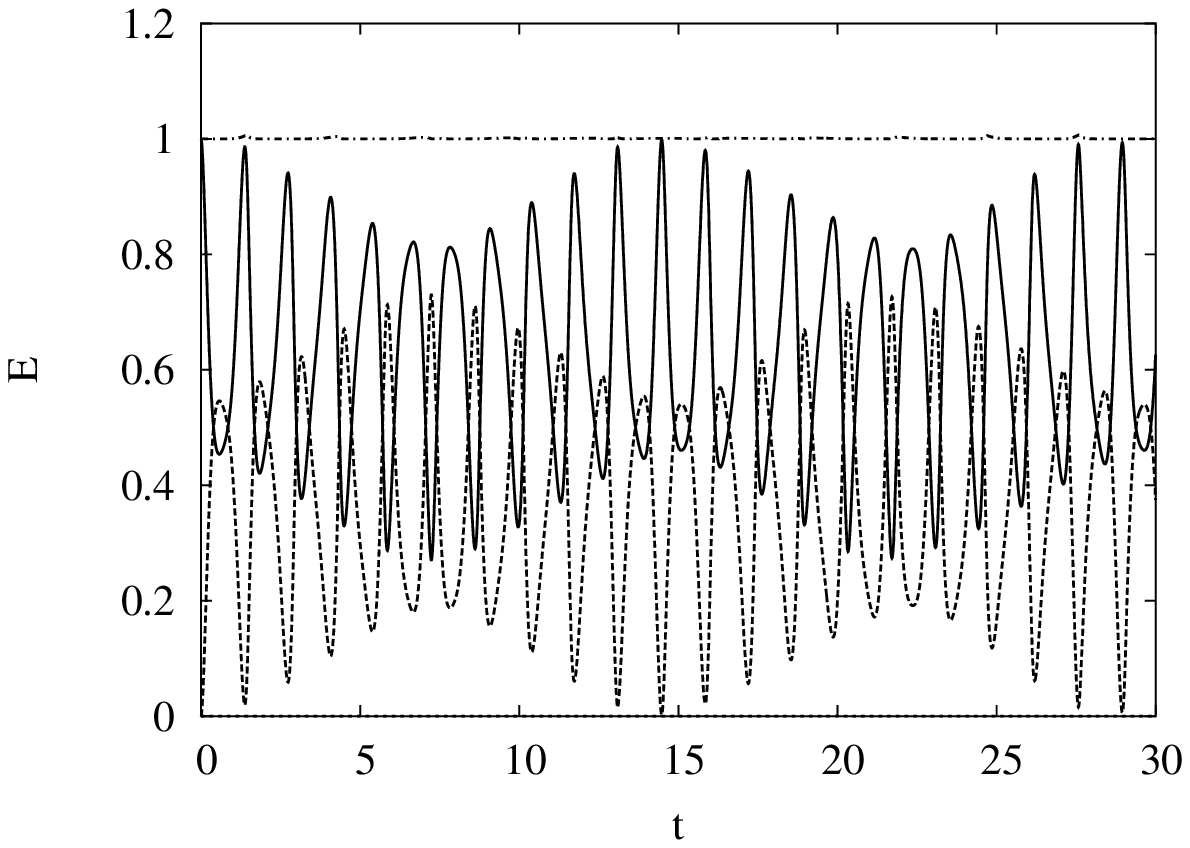}
\end{center}
\caption{Energy conservation for the $q=0.6$ case.  The curves have the
same meaning as Fig.~\ref{fig:q3e}.}
\label{fig:q6e}
\end{figure}

\section{Results}\label{sec:results}

In this paper, we calculate the timescales of the 
damping of differential rotation by magnetic fields.  We have
assumed a spherically symmetric star, having an applied (by hand)
internal dipole magnetic field (that is divergence free everywhere
and, in addition curl free everywhere except the centre).  
The initial differential rotation law is chosen (Shapiro 2000)
to be given by Eq.(\ref{eq:omega0}) and initially, toroidal magnetic 
fields are assumed to be absent.  For such a  magnetic field configuration
and initial rotation law, we solve the azimuthal components 
of the Navier--Stokes' equation and the Maxwell magnetic induction 
equation.  For computational and physical convenience, we choose to
solve these equations in a modified Magnetic Flux Co--ordinate system.
This facilitates us to solve a series of 1--D equations, parametrised
by a stream function.
Using a staggered Leap--frog scheme,
we difference the system of equations (as mentioned in Appendix) on
a uniform $q$ and $s$ grid.  In this section, we provide the results
of our computations.  For illustrative purposes, we have considered 
only 11 field lines (consistent with Figs.~\ref{fig:srho} and
\ref{fig:comp_dom}) and display results
for 3 field lines; in addition, we also display results for the surface
and the equator (essentially one point for each field line) -- for these, 
we have had to use 201 field lines for the purpose of clarity in the dynamics.

\begin{figure}
[hbp]
\begin{center}
\vspace{0.5cm}
\hspace{-0.5cm}
\includegraphics[width=0.5\textwidth]{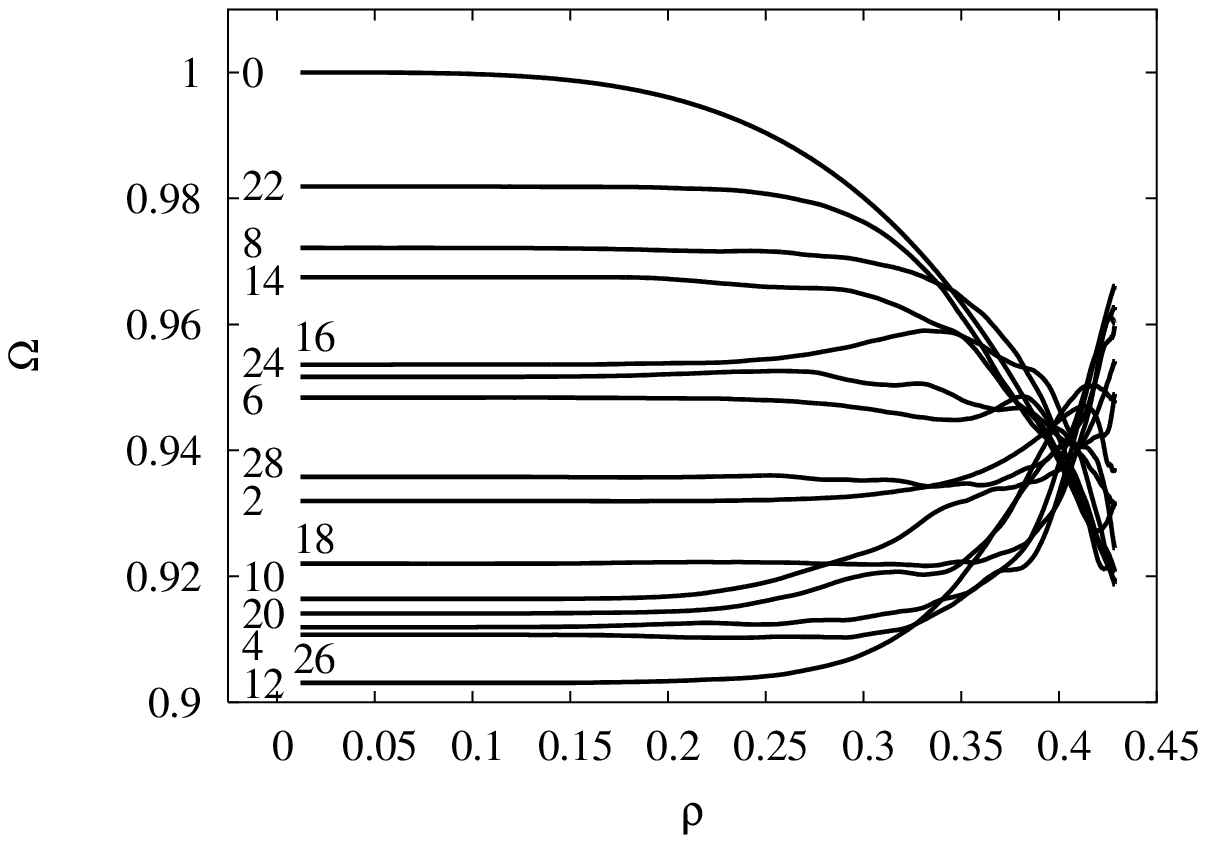}
\includegraphics[width=0.5\textwidth]{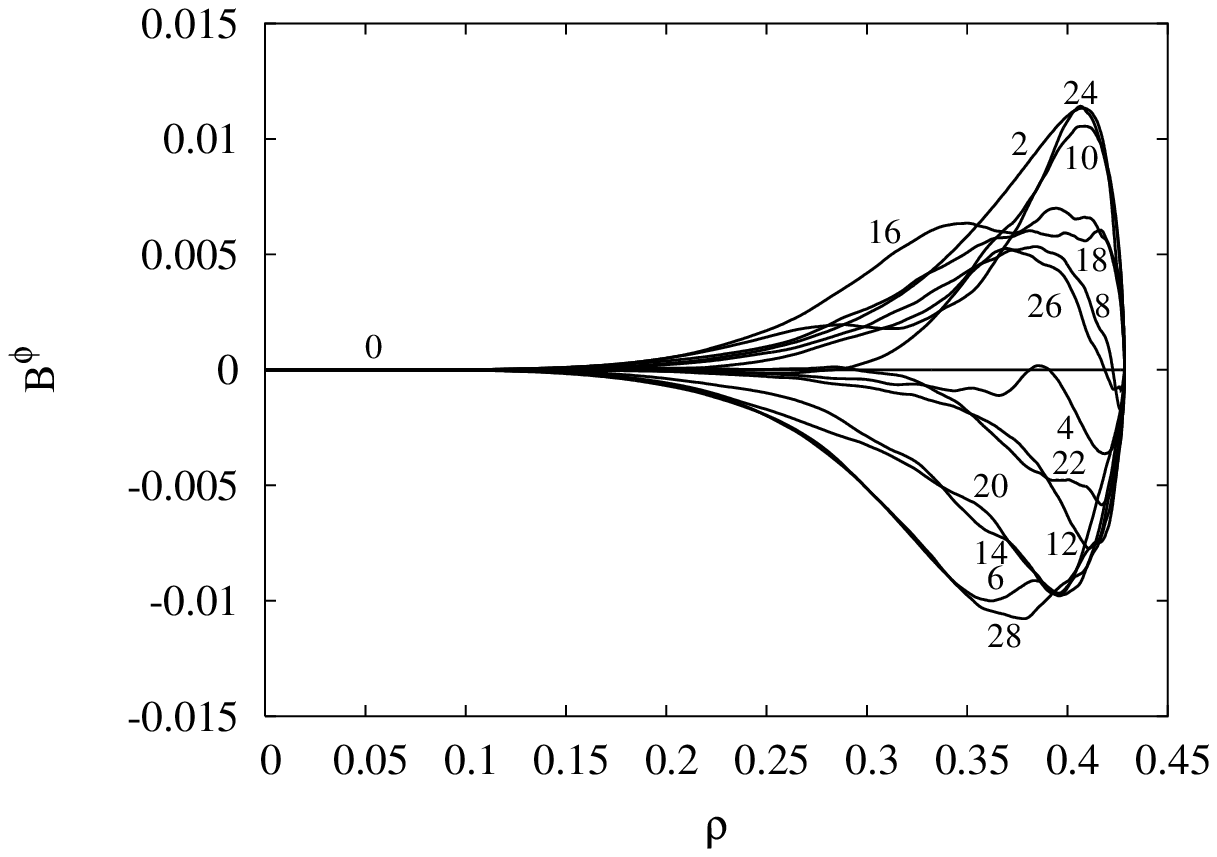}
\end{center}
\caption{$\Omega$ and $B^{\varphi}$ variation for the $q=0.8$ case.
The curves have the same meaning as for Fig. \ref{fig:q3} and
dynamics almost identical to that for $q=0.6$.}
\label{fig:q8}
\end{figure}

\subsection{Checks on the code: simulations for a monopole field configuration}
We first checked the code for a satisfactory reproduction of the 1--D 
results reported in Shapiro (2000).  A trivial identification of 
($\psi \rightarrow z$, $\varphi \rightarrow \varphi$, $\rho \rightarrow \rho$)
gives us the monopolar magnetic field.  In the above
identification, the azimuthal Eqs. (\ref{eq:mmfc1}) and (\ref{eq:mmfc2})
reduce to those in Shapiro (2000) and as is mentioned therein, 
the energy conservation is an indicator of the accuracy of the numerical 
scheme.  In Fig.~\ref{fig:conv0} we display the convergence tests for
the vacuum exterior case using
resolutions corresponding to 51, 101, 201, 401, 801 and 1601 grid points
on a single magnetic field line.  The top left panel has the minimum values
of $E_{\rm rot}$ (scaled to the value of the initial $E_{\rm rot}$).  
The straight line located at $0.513$ is the analytically 
obtained value for this quantity.  The upper and lower limit on the
y--axis correspond to a $10$\% variation from this mean value.  The top
right panel displays the maximum value for the variations in $E_{\rm mag}$
(again scaled to the initial $E_{\rm rot}$).  The analytical estimate
of $0.487$ is represented by the straight line located at the mid--point
of the y--axis in the plot displaying a maximum variation of $1$\% in
the limits.  The bottom left panel displays the $\Omega$ profile at
time $t=29$ a time where energy deviates substantially for the lowest
resolution of 51 points.  Understandably (from Eq. \ref{eq:erot}), the 
corresponding
deviations in $\Omega$ from the high resolution case is less significant
than for $E_{\rm rot}$.  The bottom right panel displays the $B^{\varphi}$
profile at $t=29$.  The ratio of the exterior density to the interior:
 $\epsilon_{\rm ex}/\epsilon = 0.2$.
\begin{figure}
[htp]
\begin{center}
\vspace{0.5cm}
\hspace{-0.5cm}
\includegraphics[width=0.5\textwidth]{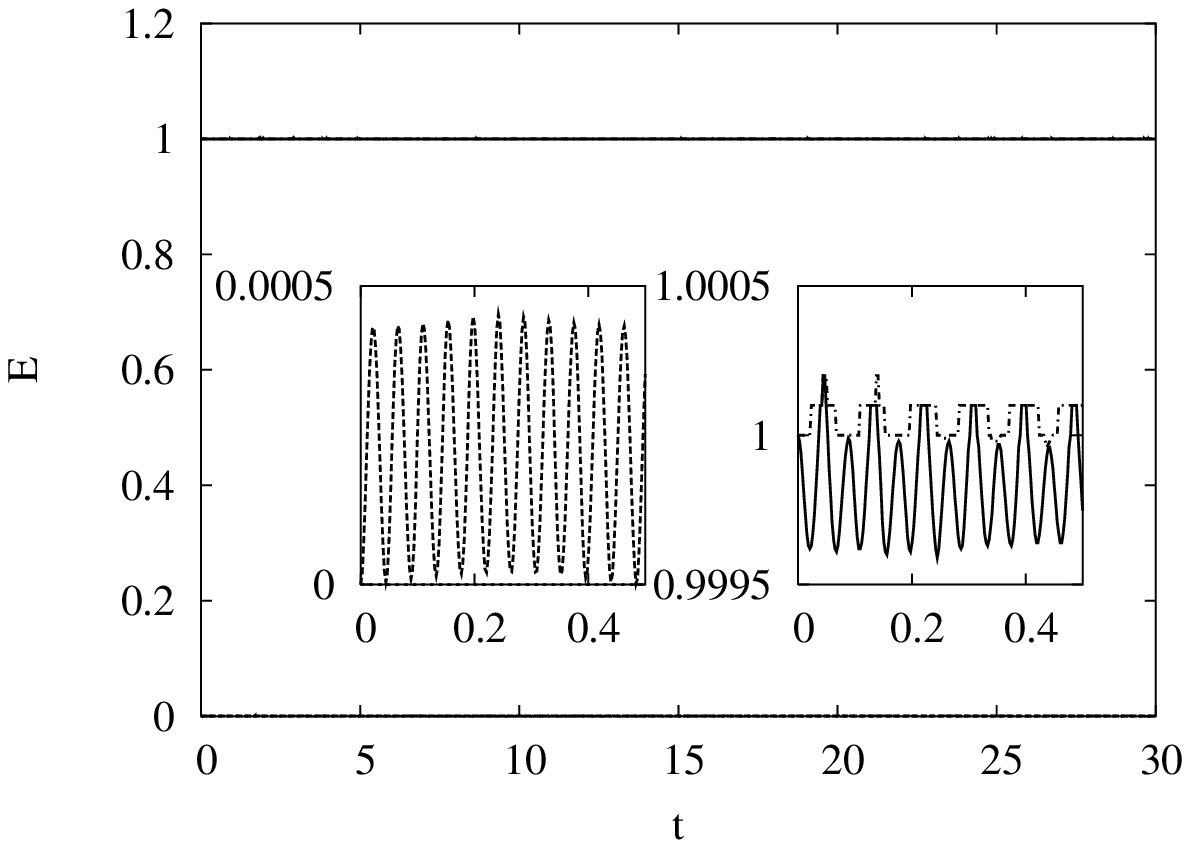}
\end{center}
\caption{Energy conservation for the $q=0.8$ case. The curves have the
same meaning as Fig.~\ref{fig:q6e}.}
\label{fig:q8e}
\end{figure}

\subsection{Simulations for the dipole field configuration}
We discuss the results for three field lines: one that is `open' 
another that is `closed' and yet another that is `marginally closed'
which we term {\it equatorial}, the latter two differing from the former
through the boundary conditions as explained in section (\ref{sec:bc}) and 
easily
deduced from Fig.~\ref{fig:comp_dom}. In particular out of the 11 field lines 
that we consider here, we discuss the results for $q=0.3$ (open), $q=0.6$ (closed) 
and $q=0.8$ (closed).

\begin{figure}
[hbp]
\begin{center}
\vspace{0.5cm}
\hspace{-0.5cm}
\includegraphics[width=0.5\textwidth]{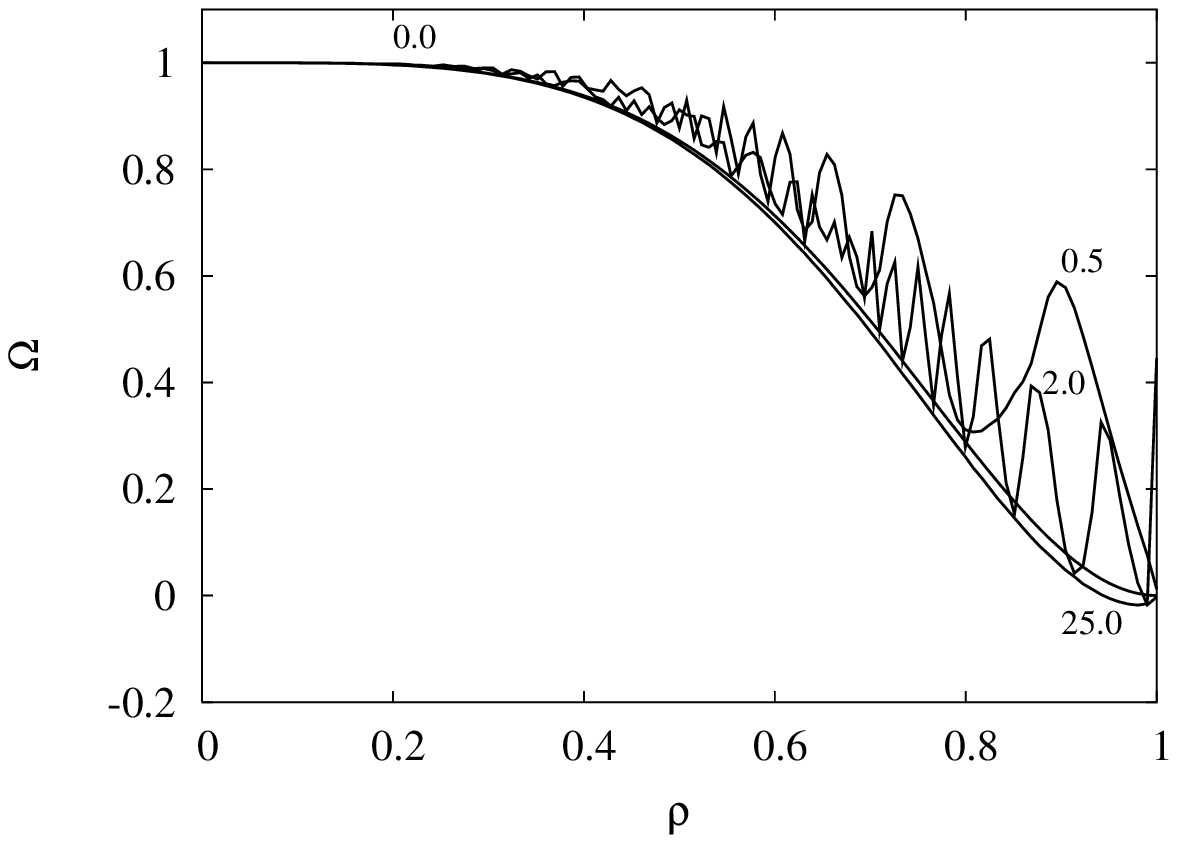}
\includegraphics[width=0.5\textwidth]{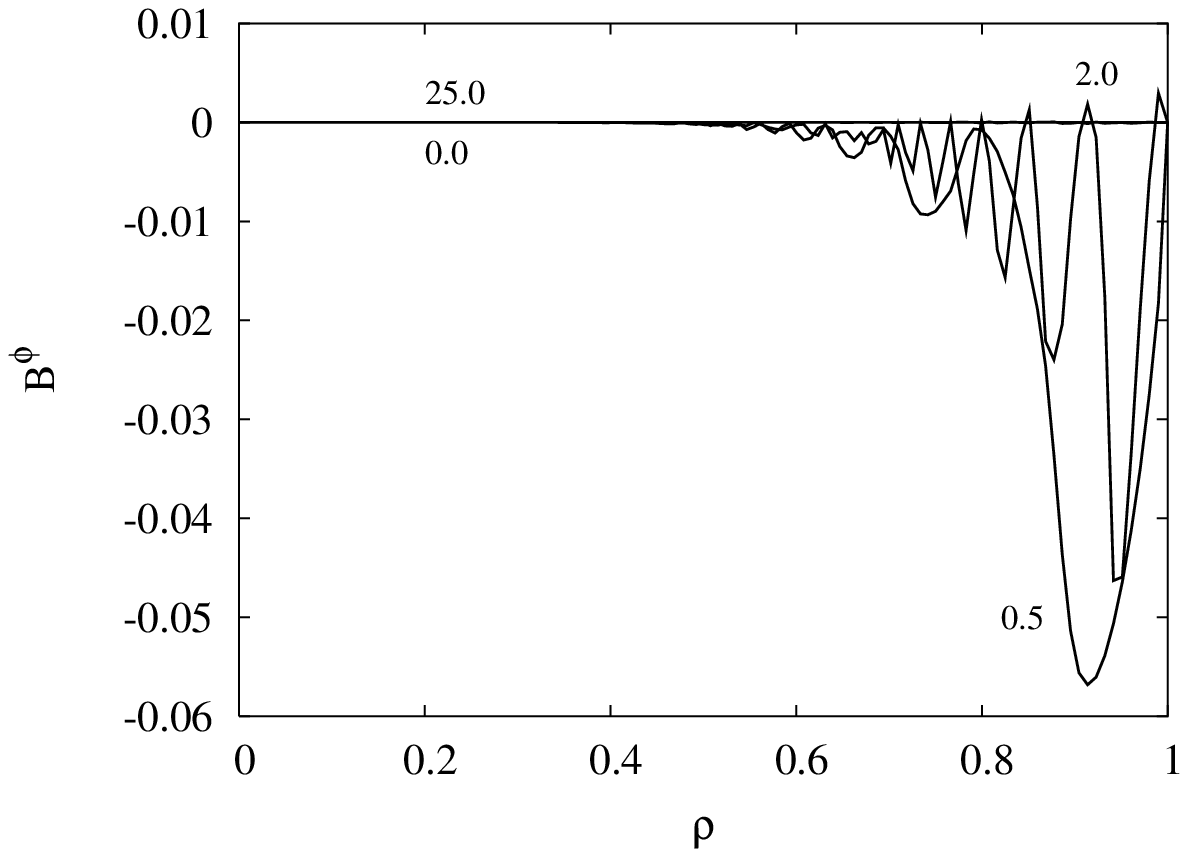}
\end{center}
\caption{$\Omega$ and $B^{\varphi}$ variation at the surface. The curves
have the same meaning as for Fig.~\ref{fig:q3}.  Structures form but
do not build up due to the radiating away of the Poynting Flux.  However,
the dynamics at the equatorial surface (last point) do not cease due to 
this being a `closed' field line.}
\label{fig:sur}
\end{figure}

\subsubsection{Results for the open field line ($q = 0.3$)}
The dynamics of the variation of the rotation and the toroidal magnetic
field are as shown in Fig.~\ref{fig:q3}. We have used
851 points along the field line direction. As can be seen, 
the torque between the stellar surface and the ambient plasma take away 
angular momentum from the star so that the rotation rate along the field line 
becomes uniform (Fig.~\ref{fig:q3} lower panel).  A plot of the energy and 
angular momentum variation over the dimensionless time $t$ is depicted in 
Fig.~\ref{fig:q3e}.  Unlike for the monopolar
case, the assumed ratio of density does not lead to retrograde rotation --
this is because the curvature of the magnetic field line produces a 
lesser torque.

\begin{figure}
[htp]
\begin{center}
\vspace{0.5cm}
\hspace{-0.5cm}
\includegraphics[width=0.5\textwidth]{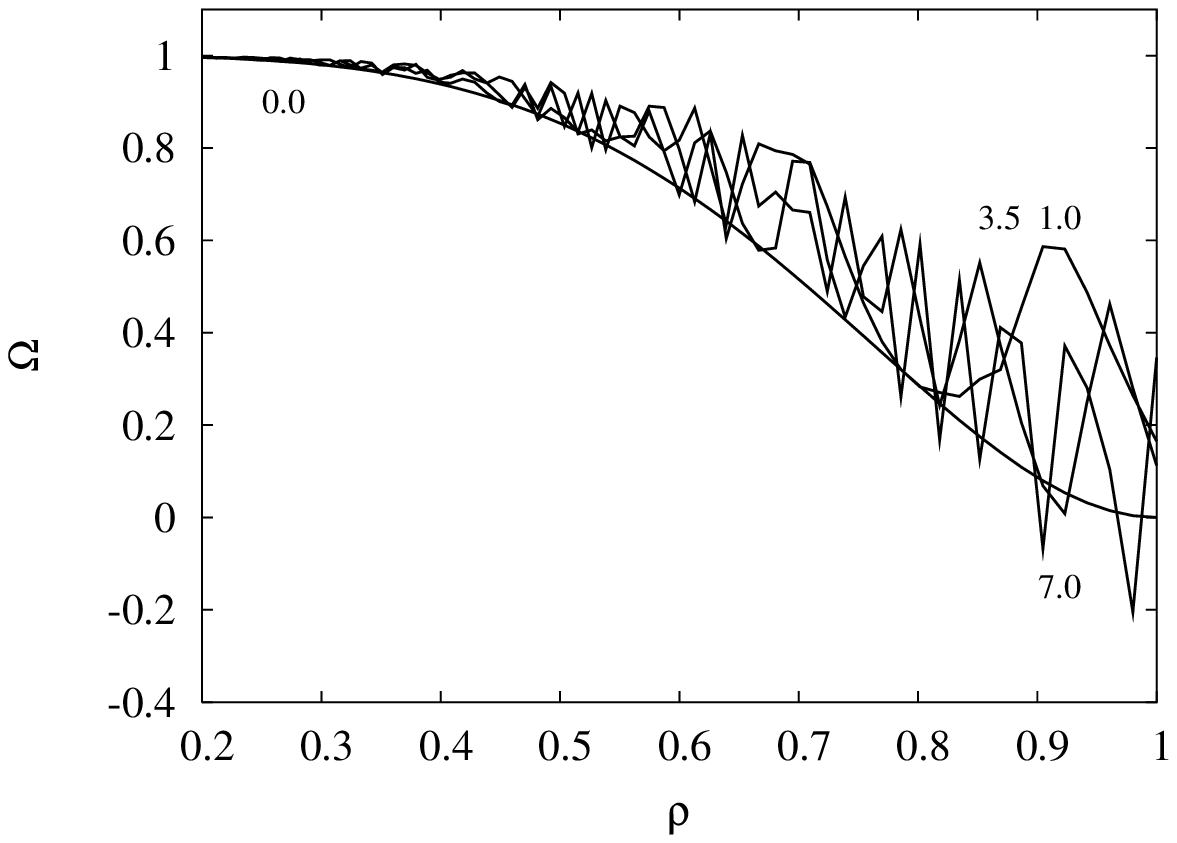}
\end{center}
\caption{$\Omega$ variation at the equator; $B^{\varphi}$ is zero. 
Unlike the case for the surface, structures build up over time to a 
point where `phase--mixing' become important.}
\label{fig:eq}
\end{figure}

\subsubsection{Results for the closed field line ($q =0.6$)}
The dynamics of the variation of the rotation and the toroidal magnetic
field are as shown in Fig.~\ref{fig:q6}.  For this
field line, we have used 1401 points in the line direction.  Since
this is a `closed' field line and there is no dissipation in the
system, the oscillations are sustained.
At late times, the toroidal magnetic field profiles develop
high gradients at the surface.  This is due to $\partial \Omega/\partial s$
not remaining zero at the surface at later times (as a result of the
singularity of $\partial z/\partial \rho$ at the equator).
The energy conservation, however, still
holds (the small variation from conserved value is due to the 
truncation error induced due to insufficient grid points).  A key
difference in the energy variation is the existence of a low frequency
component (in comparison with the 1--D counterpart) -- 
this difference is due to the fact that in the 1--D case, the substitution
of the induction equation into the Navier--Stokes equation yields
a wave equation (Shapiro 2000) in cylindrical coordinates, however,
in our general form, such a substitution results in a wave equation
with an advective component.  The low frequency component in the
energy variation with time, represents the oscillation of the 
point of inflection in our differential rotation law due to this
advection.  

\subsubsection{Results for the closed field line ($q =0.8$)}
The dynamics (as earlier) is depicted in Fig.~\ref{fig:q8}.  
The energy variation with time is shown in 
Fig.~\ref{fig:q8e}.  We have used 851 data points.  The deviation of
the total energy for the value of $1$ is due to truncation errors.

\subsection{Variation of $\Omega$ and $B^{\varphi}$ at the surface}
We have seen from Figs.~\ref{fig:q3}, \ref{fig:q6} and \ref{fig:q8} that the 
dynamics of magnetic field and rotation rate variation with time along a
given field line is smooth.  
However, if we use real (spherical/cylindrical) coordinates, the picture is 
different -- since the oscillation along a field line is uncorrelated with 
that of an immediate neighbour, beyond a certain time, the oscillations get 
out of phase as can be seen in Fig.~\ref{fig:sur} along the surface of the 
star.  Structures form, however
since Poynting flux is carried away from the star, these settle down at
late times to equilibrium values in off--equatorial regions.  Since at
the equator, there are no toroidal magnetic fields generated, the oscillation
is perpetual (unless diffusion or dissipation comes into play) -- this
fact, therefore, underlines the importance of taking into account 
diffusion to realistically model the system (Spruit~1999, 
Liu~\&~Shapiro~2004).

\subsection{Variation of $\Omega$ in the equatorial plane}
Fig.~\ref{fig:eq} illustrates the dynamical behaviour along the equator.  The 
structures
keep building up at late times here and it is important to consider
diffusion or viscous dissipation so that the neighbouring lines `talk' to
each other.

\section{Discussion}
\label{sec:discussion}
We have computed the dynamical evolution of rotation and toroidal
magnetic fields for a spherical star composed of an incompressible
homogeneous, differentially rotating fluid having an internal time
independent dipolar magnetic field.  
Given these assumption one can expect the following physical 
processes to take place:
At time $t=0$, within the spherically symmetric differentially rotating 
star, a dipolar magnetic field is switched on.  From Ferraro (1937) law of 
isorotation, we know that the lowest energy state of the system is when 
the rotation rate is constant along the magnetic field lines (provided
there is dissipation in the system).  For this
to happen, the fluid pulls the poloidal magnetic field lines, producing 
toroidal magnetic fields.  In turn the toroidal magnetic field acts back 
on the plasma.  In the absence of losses from the system, this leads to 
sustained oscillations.
For an internal dipolar magnetic field, exchange between rotational and 
toroidal magnetic field energies will take place along the `closed' field 
lines (fig) at characteristic frequencies.  Along the open field lines, 
on the other hand, there will exist non--zero toroidal surface magnetic 
fields, that will be radiated away (or go into spinning up the ambient 
plasma) leading to losses from the system.  This implies that after 
several Alf\'{v}en time--scales, Ferraro law of isorotation will hold 
on the open field lines.  

We have shown here that for the `open' field lines, rigid
rotation is achieved (in accordance with Ferraro's law) through the loss of 
Poynting flux to an ambient plasma,
while the `closed' field line execute torsional oscillation through
backreaction between toroidal magnetic and rotational kinetic energies.

Although a first glance at Eq. {\ref{eq:mmfc2}} indicates that the right hand 
side becomes $\infty$ at the centre, it can be seen from a more careful 
observation and the acknowledgement of the fact that these equations are 
for a given magnetic field line, that the quantity within parenthesis will 
be $\Omega$ 
and from our boundary conditions 
$\partial \Omega/\partial s$ becomes zero at the centre.
The upshot of this is that for different field lines, $\Omega$ possesses
different values at the centre -- a manifestation of the singularity in 
the poloidal magnetic field and $\psi$.

An important result (and verification of the results of Liu \& Shapiro
2004) is the relative difference in phases of oscillation of nearby points
at $2$--$3$ Alf\'{v}en timescales when one samples a radial or polar zone
(Figs.~\ref{fig:sur} and \ref{fig:eq}):  implying large phase deviation 
at late times (Spruit 1999).  Such a behaviour is a result of our
restrictive assumptions of neglecting diffusion and evolution
of the poloidal field.  Another point that the figures emphasise is the 
probable 
inefficacy of using pure spherical or cylindrical coordinates for analysing 
the dynamics of such a system in the absence of viscosity.  

The simplistic assumption of an internal dipolar magnetic field with `closed' 
field lines is used as an illustrative example -- the high restrictiveness 
of the sustenance of oscillation in such a system tests the efficacy of
the code and the formalism in general.  Hence, for the purposes of this
paper, the origin and sustenance of such a field in matter is taken
for granted.  The formalism and the code amply demonstrate tenacity
while the veracity of the numbers is ignored over obtaining the qualitative 
behaviour.
The key aspect of the dynamics (when there is dissipation and according to 
Ferraro's law) is that it is all about matching two 
functions (through dissipation and readjustment of the angular momentum in 
the system): one that is constant along cylinders (the initial differential 
rotation rate) to another that follows the dipolar field.

The evolution of the collapsing core of the progenitor star to pulsars 
is characterised by strong convection and turbulence.  The convection is 
driven by neutrino transport and ceases when the star becomes transparent 
to neutrinos -- an $\alpha$--$\omega$ dynamo operates in this regime 
(Thompson \& Duncan 1993; Duncan \& Thompson 1992) amplifying the primordial
magnetic fields to saturation levels $B_{\rm sat} \sim 10^{14}$~G.
We dynamically place our system at a point after the cessation of convection
and with a predominantly poloidal nature to the magnetic field -- i.e.
at about $t$\greq$30$~s after collapse.  
The next leg in the journey to computing equilibrium models of rotating
magnetised stars  will be the incorporation of less restrictive magnetic 
field geometries and the effects of dissipation.

\begin{acknowledgements}
The author thanks 
H.M. Antia 
and 
Kandaswamy Subramaniam 
for discussions.  
Thanks are also due to Bobo Ahmedov, 
Ranjeev Misra,
Koji Uryu, 
and 
Shin Yoshida, 
for a critical reading of the manuscript and stimulating
discussions. 
Luciano Rezzolla and John Miller are thanked for their suggestion of this 
project.  This work was supported by visiting fellowships at TIFR and IUCAA.  
The author thanks Alak Ray, Ajit Kembhavi and the deans and directors of TIFR
and IUCAA for kind hospitality.  The computations were performed on the 
``Compaq Tru 64'' clusters at TIFR and IUCAA computer centres.
\end{acknowledgements}

\appendix 

\section{Numerical Scheme}

In this section we describe the numerical scheme that we employ.  The
equations that we solve are of the type:

\begin{eqnarray}
\frac{\partial X}{\partial t} & = & {\cal A} 
\frac{\partial} {\partial s} ({\cal B} Y)
\label{eq:appone}
\\
\frac{\partial Y}{\partial t} & = & {\cal C} 
\frac{\partial} {\partial s} ({\cal D} X)
\label{eq:apptwo}
\end{eqnarray}

We assume $X$ and $Y$ to be functions of $t$, $s$ and $q$ (dependence
on $q$ is implicit as the eqns. (\ref{eq:appone}) and (\ref{eq:apptwo})
are for constant values $q$).
When we finite difference the above equations, we define $X$ and $Y$ on
alternate grid points so as to yield a staggered mesh (and hence a 
second order treatment for space).  Typically, we demand that $X$ be
defined on `half' grid points (cell boundaries) and $Y$ on
`full' grid points (cell centres). Accordingly, the dependent variables
will carry the tag:

\begin{eqnarray}
X(t,s,q) & = & X^{n}_{i+1/2,j},~~X^{n}_{i+3/2,j} ~~ ...
~~(\displaystyle i~~\epsilon~~[1,N-1]) \\
Y(t,s,q) & = & Y^{n}_{i,j},~~Y^{n}_{i+1,j} ~~ ...
~~(\displaystyle i~~\epsilon~~[1,N])
\end{eqnarray}
where $N$ is the total no. of grid points along a given magnetic field
line (i.e. a given $q$).
Since we're working on constant $q$ (and hence constant $j$) lines,
for the sake of easy representation here, we shall drop the subscript $j$ 
from the variables (nevertheless remembering that the variables are indeed
three dimensional). Thus, for a given $q$:
\begin{eqnarray}
X(t,s) & = & X^{n}_{i+1/2},~~X^{n}_{i+3/2} ~~ ...
~~(\displaystyle i~~\epsilon~~[1,N-1]) \\
Y(t,s) & = & Y^{n}_{i},~~Y^{n}_{i+1} ~~ ...
~~(\displaystyle i~~\epsilon~~[1,N])
\end{eqnarray}


%


We discretise Eqs. (\ref{eq:appone}) and (\ref{eq:apptwo})
in staggered leap--frog scheme as mentioned below:
\begin{eqnarray}
\frac{X^{\small n+1}_{\small i+1/2} - 
X^{\small n}_{\small i+1/2}}{\Delta t} & = &
{\cal A}_{\small i+1/2} 
\left(\frac{{\cal B}_{\small i+1} Y^{\small n}_{i+1} - 
{\cal B}_{\small i} Y^{\small n}_{i}}{\Delta s}\right) 
\label{eq:scheme2.1}
\\
\frac{Y^{\small n+1}_{\small i} - Y^{\small n}_{\small i}}{\Delta t} & = &
{\cal C}_{\small i} 
\left(\frac{{\cal D}_{\small i+1/2} X^{\small n}_{\small i+1/2} - 
{\cal D}_{\small i-1/2} X^{\small n}_{\small i-1/2}}{\Delta s}\right) 
\label{eq:scheme2.2}
\end{eqnarray}

{}
\end{document}